\documentclass[prd,english,preprintnumbers,amsmath,amssymb,nofootinbib,superscriptaddress,twocolumn]{revtex4-1}

\pdfoutput=1

\usepackage{mathtools}
\usepackage{amsfonts}
\usepackage{mathrsfs}
\usepackage{bbm}
\usepackage{slashed}
\usepackage{appendix}

\usepackage{graphicx}
\usepackage[dvipsnames]{xcolor}
\usepackage{array}

\usepackage{xspace}
\usepackage{siunitx}
\usepackage{hyperref}

\usepackage{xifthen}
\usepackage{dsfont}
\usepackage{appendix}
\usepackage{enumitem}
\usepackage{booktabs}
\usepackage{units}
\usepackage{appendix} 

\usepackage[latin1]{inputenc}
\usepackage[dvipsnames]{xcolor}
\usepackage{array}

\setkeys{Gin}{width=0.48\textwidth}
\graphicspath{
	{./}
}

\graphicspath{
	{./}
	{../}
}

\def\0#1#2{\frac{#1}{#2}}

\def\s0#1#2{\mbox{\small{$ \frac{#1}{#2} $}}}

\newcommand{\I}{\mathrm{i}}
\newcommand{\be}{\begin{eqnarray}}
\newcommand{\ee}{\end{eqnarray}}

\newcommand{\nn}{\nonumber }

\newcommand{\beq}{\begin{equation}}
\newcommand{\eeq}{\end{equation}}
\newcommand{\bea}{\begin{eqnarray}}
\newcommand{\eea}{\end{eqnarray}}
\newcommand{\Nc}{N_{\rm c}}
\newcommand{\Nf}{N_{\rm f}}

\newcommand{\SpPmAdj}{{(S+P)_{-}^\mathrm{adj}}}
\newcommand{\Csc}{{\mathrm{csc}}}

\newcommand{\VmAPar}{{(V-A)_{\parallel}}}
\newcommand{\VmAPer}{{(V-A)_{\perp}}}
\newcommand{\VpAPar}{{(V+A)_{\parallel}}}
\newcommand{\VpAPer}{{(V+A)_{\perp}}}

\newcommand{\VmAPerAdj}{{(V-A)_{\perp}^{\mathrm{adj}}}}
\newcommand{\VpAParAdj}{{(V+A)_{\parallel}^{\mathrm{adj}}}}

\def\0#1#2{\frac{#1}{#2}}

\def\eq#1{\eqref{#1}}

\newcommand{\gettitle}{Chiral and effective $U(1)_{\rm A}$ symmetry restoration in QCD}

\hypersetup{
	colorlinks,
	linkcolor={blue!75!black},
	citecolor={blue!75!black},
	urlcolor={blue!75!black},
	pdftitle={\gettitle},
	pdfauthor={Braun, Leonhardt, Rosenblueh},
	pdfkeywords={Renormalization group} {Cutoff}
	{Correlations functions} {Initial conditions},
	bookmarksopen=true,
	bookmarksopenlevel=2,
	bookmarksnumbered=true
}

\usepackage{babel}
\makeatother
\begin{document}

\title{Chiral and effective $U(1)_{\rm A}$ symmetry restoration in QCD}

\author{Jens Braun}
\affiliation{Institut f\"ur Kernphysik (Theoriezentrum), Technische Universit\"at Darmstadt, 
D-64289 Darmstadt, Germany}
\affiliation{ExtreMe Matter Institute EMMI, GSI, Planckstra{\ss}e 1, D-64291 Darmstadt, Germany}
\author{Marc Leonhardt} 
\affiliation{Institut f\"ur Kernphysik (Theoriezentrum), Technische Universit\"at Darmstadt, 
D-64289 Darmstadt, Germany}
\author{Jan M. Pawlowski}
\affiliation{Institut f\"ur Theoretische Physik, Universit\"at Heidelberg, Philosophenweg 16, 69120 Heidelberg, Germany}
\affiliation{ExtreMe Matter Institute EMMI, GSI, Planckstra{\ss}e 1, D-64291 Darmstadt, Germany}
\author{Daniel Rosenbl\"uh} 
\affiliation{Institut f\"ur Kernphysik (Theoriezentrum), Technische Universit\"at Darmstadt, 
D-64289 Darmstadt, Germany}

\begin{abstract}
The nature and location of the QCD phase transition close to the chiral limit restricts the phase structure of QCD with physical pion masses at non-vanishing density. At small pion masses, explicit $U(1)_{\rm A}$-breaking, as induced by a non-trivial topological density, is of eminent importance. It triggers the 't Hooft interactions and also manifests itself in the interplay of four-quark interactions at low momentum scales. In the present work, we perform a Fierz-complete analysis of the emergence of four-quark interactions from the QCD dynamics at finite temperature, subject to a given 't Hooft coupling at large momentum scales. The variation of the latter allows us to test the robustness of our findings. Taking an estimate of the effect of the topological running of the 't Hooft coupling into account, our analysis suggests that the chiral transition in QCD with two massless quark flavours falls into the $O(4)$ universality class. 
\end{abstract}

\maketitle

%
\section{Introduction}
The dynamics of Quantum Chromodynamics (QCD) close to the chiral limit constrains the QCD phase structure with physical pion masses at non-vanishing density. This has led to a rekindled interest in the nature of the chiral transition at vanishing baryon density in recent years. While both, functional QCD (e.g., Refs.~\cite{Fu:2019hdw, Braun:2019aow, Fischer:2018sdj, Isserstedt:2019pgx, Gao:2020qsj, Gao:2020fbl}) and lattice QCD approaches (e.g., Refs.~\cite{Aoki:2006we,Aoki:2006br, Bonati:2018nut, Borsanyi:2018grb, Bazavov:2018mes, Guenther:2018flo, Ding:2019prx}), exhibit a crossover for physical quark masses,  the situation close to the chiral limit is unresolved to date, in particular for QCD with two light quark flavours~\cite{Braun:2009gm, Cossu:2013uua, Philipsen:2014rpa, Dick:2015twa, Tomiya:2016jwr, Philipsen:2016hkv, Cuteri:2017gci}. 

The order of the chiral transition and the associated universality class is known to depend crucially on the fate of the~$U(1)_{\rm A}$ axial symmetry at the critical temperature~\cite{Pisarski:1983ms, Mitter:2013fxa, Grahl:2013pba, Pelissetto:2013hqa, Fejos:2016hbp, Rennecke:2016tkm, Resch:2017vjs, Li:2019chs}. Therefore, the analysis of the effect of the axial anomaly on the QCD phase structure is a very challenging problem both for functional and lattice approaches to QCD. To be more specific, for QCD with two massless quark flavours, a second-order phase transition in the~$O(4)$-universality class is only expected if the chiral transition and effective $U(1)_{\rm A}$ restoration are sufficiently separated in temperature~\cite{Pisarski:1983ms, Grahl:2013pba, Pelissetto:2013hqa}. Whereas indications for an (almost) coincidence of the chiral transition and effective $U(1)_{\rm A}$ restoration have  been observed in lattice studies with Wilson fermions~\cite{Brandt:2016daq}, lattice studies based on the HISQ action suggest that both are sufficiently separated such that the $O(4)$ scenario appears to be favoured~\cite{Kaczmarek:2020sif}.

Besides the explicit breaking of the~$U(1)_{\rm A}$ symmetry, investigations of the nature of the chiral transition in two-flavour QCD are further complicated by the fact that it may also be very sensitive to the quark mass. This can be seen by embedding QCD with two quark flavours into three-flavour QCD. For the latter, the chiral transition is expected to be of first order for sufficiently small masses of the three quark flavours~\cite{Pisarski:1983ms}. However, the size of this first-order region may even extend to the limit of infinitely heavy strange quarks which is associated with two-flavour QCD~\cite{Philipsen:2016hkv, Cuteri:2017gci, Cuteri:2017zcb}. Studies of low-energy effective theories indeed suggest that the phase transition may be of first order, depending on the strength of the $U(1)_{\rm A}$ breaking and the values of the quark masses, see, e.g., Refs.~\cite{Mitter:2013fxa, Rennecke:2016tkm, Resch:2017vjs, Pisarski:2019upw}.

The present work aims at an understanding of the dependence of the chiral phase transition on the strength of the anomalous $U(1)_{\rm A}$ breaking and represents an important step for more detailed future studies of the nature of the chiral transition at small quark masses based on functional approaches to QCD. In fact, as any kind of ``deformation" of a system by a variation of an external parameter allows to gain a deeper insight into the microscopic dynamics, a study of the response of QCD under the variation of the strength of anomalous $U(1)_{\rm A}$ breaking also opens up the opportunity to unravel further details of the QCD dynamics close to the phase transition.  

In the following, we focus on a study of the renormalisation group (RG) flow of gluon-induced four-quark interaction channels in a Fierz-complete setting and analyse chiral symmetry restoration in the zero-density limit. Anomalous $U(1)_{\rm A}$ symmetry breaking at large momentum scales is introduced in our study via the 't Hooft determinant~\cite{tHooft:1976rip, tHooft:1976snw, THOOFT1986357}, see Ref.~\cite{Pisarski:2019upw} for a systematic generalisation to higher instanton numbers. Within the RG approach to QCD, this has been first studied in Ref.~\cite{Pawlowski:1996ch}.  Importantly, for two quark flavours, the 't Hooft determinant leads to a $U(1)_{\rm A}$-breaking four-quark interaction channel. The value of the so-called 't Hooft coupling measuring the strength of this channel and the strong coupling at the initial RG scale are then the two only input parameters of our present study. 

In combination with the significant advances made with respect to functional first-principles studies of QCD in the past 15 years, we expect the present work to open up the opportunity for future quantitative functional studies of the nature of the QCD transition close to the chiral limit. For advances with the functional RG approach related to the current work, we refer the reader to Refs.~\cite{Braun:2007bx, Fischer:2008uz, Braun:2008pi, Braun:2009gm, Braun:2014ata, Mitter:2014wpa, Rennecke:2015eba, Cyrol:2017ewj, Cyrol:2017qkl, Fu:2019hdw, Braun:2019aow, Leonhardt:2019fua, Gao:2020fbl, Gao:2020qsj}. In full QCD RG-flows, the 't Hooft coupling will no longer be an input parameter. There, it will be generated by $U(1)_{\rm A}$-breaking fluctuations. We emphasise that the initial value of the 't Hooft coupling in the current study takes care of both, the large momentum-scale limit of $U(1)_{\rm A}$ breaking as well as the potential lack of topological fluctuations. After a more general discussion of some of these aspects in \autoref{sec:ua14f}, we analyse chiral symmetry restoration and effective $U(1)_{\rm A}$ restoration over a wide range of temperatures and values of the initial 't Hooft coupling within our framework in \autoref{sec:ua1T}. Our conclusions are then presented in \autoref{sec:conc}.


%
\section{$U(1)_{\rm A}$ symmetry and four-quark interactions}\label{sec:ua14f}
For our study of chiral and effective $U(1)_{\rm A}$ restoration at finite temperature, we use the functional renormalisation group (fRG) approach, based on the Wetterich equation~\cite{Wetterich:1992yh} for the quantum effective action~$\Gamma$. In the following subsections, we discuss the truncation of the full effective action underlying our present work, with a focus on four-quark interaction channels and aspects associated with the breaking of the $U(1)_{\rm A}$ symmetry. 

\subsection{Effective action}\label{sec:FlowG}
Within the fRG approach, the effective action~$\Gamma_k$ depends on an infrared (IR) cutoff scale~$k$ which is integrated out successively. For large cutoff scales $k$, the infrared regularised effective action is well-described by perturbation theory and finally reduces to the classical action, depending only on the UV-relevant terms. In turn, for vanishing infrared cutoff scales, $k\to 0$, the full effective action of QCD emerges. For QCD-related reviews, see Refs.~\cite{Pawlowski:2005xe, Gies:2006wv, Rosten:2010vm, Braun:2011pp, Pawlowski:2014aha, Dupuis:2020fhh}. 

In our present study we rely on the background field approach to gauge theories within background covariant gauges and employ the well-studied background field approximation, where correlation functions of the full field are identified with that of the background field. For a detailed general discussion, we refer the reader to, e.g., Ref.~\cite{Dupuis:2020fhh}. This leads to a gauge-invariant effective action. In this work, we consider a sum of the classical QCD action with running couplings for the physical case of three colours, $\Nc=3$, augmented with a Fierz-complete set of four-quark interactions, see also Ref.~\cite{Braun:2019aow}:
\begin{subequations}
	\label{eq:ApproxG}
\begin{align}\nonumber 
	\Gamma_k=  &\,\int {\rm d}^4x\left\{ 
	\frac{1}{4}\,F_{\mu\nu}^{a}F_{\mu\nu}^{a}
	+ \bar{q}\left(
	{\rm i}\slashed{\partial} + g\slashed{A} \right)q
	\right\}\\[1ex]
	& \qquad \qquad \qquad +S_{\rm gf}+ S_\textrm{gh}+ \Delta\Gamma_{4\textrm{-quark}}\,. 
\label{eq:Gamma}\end{align}
Here, $F_{\mu\nu}^a =\partial_\mu A^a_\nu-\partial_\nu A^a_\mu + g f^{abc} A_\mu^b A_\nu^c$, $a=1,...,8$, and  $\Delta\Gamma_{4\textrm{-quark}}$ stands for four-quark terms. In Eq.~\eq{eq:Gamma}, we have absorbed the wave function renormalisation of the background field into the gauge field, and $g$ is the full scale-dependent (background) QCD coupling. Note that we do not take into account the running of the wave function renormalisation of the quark fields, as it depends only mildly on the RG scale~\cite{Gies:2002hq,Braun:2008pi,Braun:2014ata, Rennecke:2015eba, Mitter:2014wpa, Cyrol:2017ewj, Fu:2019hdw}.

With respect to four-quark interactions, we restrict ourselves to the pointlike limit, see also our discussion in \autoref{sec:4q}. This entails that $\Delta\Gamma_{4\textrm{-quark}}$ can be written in a Fierz-complete sum of single local four-quark terms where the quarks and anti-quarks are contracted with tensors ${\mathcal O}_i$ forming a complete tensor basis:
\begin{align}\label{eq:four-quark} 
	\Delta\Gamma_{4\textrm{-quark}}= \sum_{i=1}^{N_\textrm{Fierz}} \Delta\Gamma_{4\textrm{-quark},i}\,,
\end{align}
where 
\begin{align}
\Delta\Gamma_{4\textrm{-quark},i} = \bar{\lambda}_{i} \int {\rm d}^4x\,\left(\bar{q}\,{\mathcal O}_i\,q\right)^2\,
\label{eq:delta4f}
\end{align}
\end{subequations}
with $k$-dependent couplings $\bar{\lambda}_{i}$. The tensor~${\mathcal O}_i$ determines the colour, flavour, and Dirac structure of the respective vertex. The details of the employed four-quark basis and the running of the respective couplings will be discussed below. Here, we only state that the four-quark interactions are generated from the fundamental quark-gluon interactions in QCD and hence carry the symmetries of the classical QCD action. Note that the classical QCD action is invariant under $SU(\Nc)\otimes SU(\Nf)_\text{L}\otimes SU(\Nf)_\text{R}\otimes U(1)_\text{V}\otimes U(1)_\text{A}$ transformations of the quark fields where~$\Nf$ specifies the number of flavours. 

\subsection{$U(1)_{\rm A}$-violating multi-quark interactions}\label{sec:UA1}
The ${U(1)_{\rm A}}$ anomaly triggers a ${U(1)_{\rm A}}$-violating multi-quark interaction term $\Delta\Gamma_{U(1)_{\rm A}}$ which is generated from a non-trivial topological density. This has been first derived from semi-classical considerations in an expansion about nontrivial topological gauge configurations~\cite{tHooft:1976rip, tHooft:1976snw, THOOFT1986357}. Within the present background-field fRG approach to QCD, the 't Hooft term has been derived in Ref.~\cite{Pawlowski:1996ch}, see also Ref.~\cite{Hamada:2020mug} for applications. 

The semi-classical approach is valid at large momentum scales. In this regime, $\Delta\Gamma_{U(1)_{\rm A}}$ assumes the form of a $(2\Nf)$-quark interaction channel which is only invariant under $SU(\Nc)\otimes SU(\Nf)_\text{L}\otimes SU(\Nf)_\text{R}\otimes U(1)_\text{V}$ transformations, i.e., it explicitly breaks the $U(1)_\text{A}$ symmetry present in the classical QCD action. Note that anomalous $U(1)_\text{A}$ breaking also generates lower order terms in the case of finite current quark masses. 

For~$\Nf=2$, the ${U(1)_{\rm A}}$-violating term $\Delta\Gamma_{U(1)_{\rm A}}$ is a four-quark interaction and can be taken as one of the four-quark tensor structures in the sum~\eq{eq:four-quark}. It reads  
\begin{align}
\Delta\Gamma_{U(1)_{\rm A}} \!=\! {\bar\lambda}_\textrm{top} \int {\rm d}^{4}x \,\Bigl(
{\text{det}} \left[ \bar{q}_i P_{\rm L} q_j \right] \!+\! {\text{det}} \left[ \bar{q}_i P_{\rm R} q_j \right]
\Bigr)\,,
\label{eq:gua1}
\end{align}
where~$P_{\text{R/L}}= \frac{1}{2}({1 \pm \gamma_5})$, the indices $i,j$ refer to the quark flavours, and the determinant is taken in flavour space. In the following, we shall often refer to the coupling~${\bar\lambda}_\textrm{top}$ associated with this interaction channel as the 't Hooft coupling. Taking into account the explicit $U(1)_{\rm A}$ breaking induced by this channel, we end up with $N_\textrm{Fierz}=10$ four-quark terms for two {\it massless} quark flavours, see \hyperref[app:2]{App.~\ref{app:2}}. For two {\it massless} and one \textit{heavy} flavour, anomalous $U(1)_\text{A}$ breaking induces a six-fermion interaction (plus lower order terms). The dimension of the corresponding Fierz-complete basis of four-quark interactions is then~$N_{\textrm{Fierz}}=32$, see \hyperref[app:2+1]{App.~\ref{app:2+1}} for details.

Note that, for large cutoff scales, the four-quark part of the effective action decays at least as $\sim 1/k^2$. The ${U(1)_{\rm A}}$-violating terms drop even faster as they come with a non-perturbative prefactor of the form $\sim e^{-8 \pi^2/g^2}$ which is induced by the topological density. The latter factor leads to an additional polynomial decay because of the logarithmic running of the strong coupling at large scales, see Ref.~\cite{Pawlowski:1996ch}. The rapid decay of the four-quark terms suggests that they can in principle be dropped at the initial cutoff scale $\Lambda$, provided it is chosen sufficiently large. Whereas this reasoning indeed applies to $U(1)_{\rm A}$-symmetric channels, it does not apply to the $U(1)_{\rm A}$-violating term~$\Delta\Gamma_{U(1)_{\rm A}}$, although it decays even faster as a function of~$k$. Indeed, a finite initial condition for the latter term is required to actually trigger $U(1)_{\rm A}$ breaking. To be specific, if we set $\Delta\Gamma_{U(1)_{\rm A}}$ as well as all other four-quark couplings to zero at the initial scale~$\Lambda$, then the $U(1)_{\rm A}$ symmetry is intact at this scale. Interactions of the form of~$\Delta\Gamma_{U(1)_{\rm A}}$ are then still generated dynamically in the flow. However, the $U(1)_{\rm A}$-symmetry breaking potentially induced by this term is always compensated by other simultaneously generated $U(1)_{\rm A}$-violating interaction channels such that the $U(1)_{\rm A}$ symmetry remains intact in the RG flow, see also \autoref{sec:sr_intro} below.

Let us now estimate the size of the topological contributions~$\sim e^{-8 \pi^2/g^2}$ to the running of the four-quark couplings. To this end, we exploit the findings in Ref.~\cite{Pawlowski:1996ch}, where the purely topological running of $\bar\lambda_\textrm{top}$ has been studied based on the dilute gas approximation. In recent lattice studies~\cite{Borsanyi:2015cka, Petreczky:2016vrs, Jahn:2020oqf}, the latter approximation has been found to be a good approximation at high temperatures. In the following, we shall also expect this approximation to hold true in the perturbative and semi-perturbative regime associated with scales  $k\gtrsim 2\,\text{GeV}$. Below this scale, the topological contributions to the running of the 't Hooft coupling~${\bar\lambda}_\textrm{top}$ are finally overcome by quark self-interactions associated with the generated terms, see Ref.~\cite{Pawlowski:1996ch} for a first discussion. This effect is related and very similar to the flow of the other four-quark interactions: they are generated and driven by two-gluon exchange diagrams over a wide momentum range. However, for~$k\lesssim 2\,\text{GeV}$, the quark self-interactions eventually start to dominate the RG flow, see also \autoref{sec:ua1T}. 
\begin{figure}[t]
	\begin{center}
		\includegraphics[width=1\linewidth]{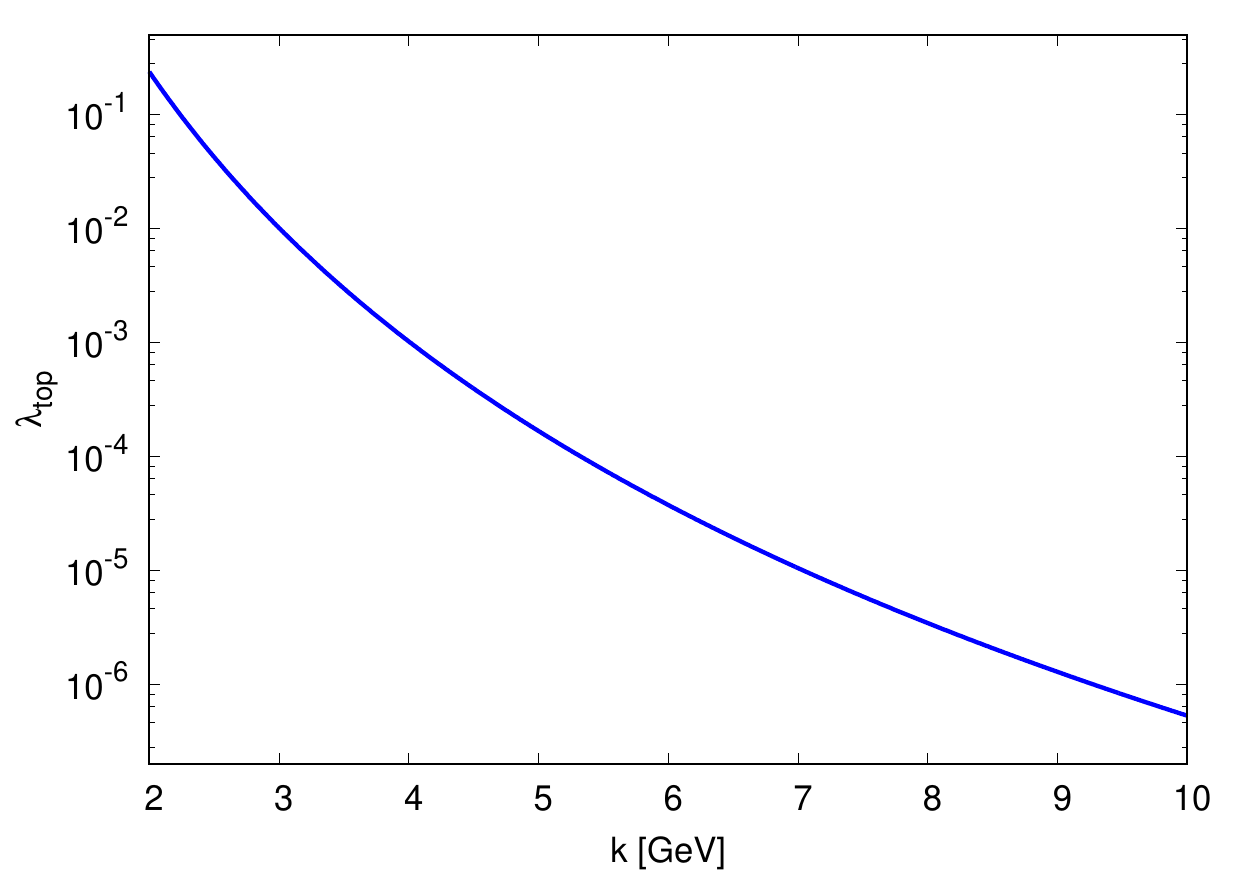}
	\end{center}
	\vspace*{-0.5cm}
	\caption{Scale dependence of the dimensionless 't Hooft coupling as induced by fluctuations around instantons in the approximation from Ref.~\cite{Pawlowski:1996ch}. The increase towards lower scales is mainly driven by the factor~$\sim{\rm e}^{-8\pi^2/g^2}$ because of the increase of the strong coupling. The integral over the size of the instanton is infrared regularised by both the cutoff term and the gluon condensate with $g^2\langle F^a_{\mu\nu} F^a_{\mu\nu}\rangle/(4\pi^2) \sim  0.012\,\textrm{ GeV}^4$.}
	\label{fig:lambdatop_top}
\end{figure}

For the discussion of the strength of the 't Hooft channel as well as of the other four-quark channels below, it is convenient to consider dimensionless couplings~$\lambda_i$:
\begin{align}
	\label{eq:dimless4q}
	\lambda_i := \bar{\lambda}_i k^2  \,.
\end{align} 
Within the approximations made in Ref.~\cite{Pawlowski:1996ch}, including an action term proportional to the gluon condensate that serves as a physical infrared cutoff, the purely topological contributions then lead to the coupling strength~$\lambda_\textrm{top}\sim 10^{-1}$ at~$k=2\,\text{GeV}$. For increasing~$k$, $\lambda_\textrm{top}$ is found to decrease monotonically. At~$k=10\,\text{GeV}$, which we use as initial RG scale below, we end up with $\lambda_\textrm{top}\sim 10^{-7}$, see \autoref{fig:lambdatop_top} for an illustration of the scale dependence of~$\lambda_\textrm{top}$.

In the present work, we do not take into account topological contributions to the RG flow of the 't Hooft coupling~$\lambda_{\text{top}}$. Instead, we phenomenologically account for these contributions with the initial condition for $\lambda_{\textrm{top}}$ which therefore serves as ``external" control parameter. Given the results for the topological running depicted in \autoref{fig:lambdatop_top}, we mostly restrict our present analysis to the following range of initial values of the 't Hooft coupling: 
\begin{align}\label{eq:lambdatoprange}
\lambda_{\textrm{top}}(k=\Lambda) \in [10^{-1} , 10^{-7}]\,.
\end{align}
We expect this range to also include the phenomenologically relevant value. In any case, we shall see below that the IR dynamics decouples from the initial condition over this wide range of values in the zero-temperature limit. On the other hand, it is well-known that the finite-temperature behaviour reflects the scale dependence of the couplings which we will study in detail below.  

 \subsection{Fierz-complete four-quark interactions}\label{sec:4q}
While the $U(1)_\textrm{A}$-violating four-quark interactions are generated from topological fluctuations, the other tensor channels in Eq.~\eq{eq:four-quark} are compatible with the symmetries of the classical QCD action. The couplings $\lambda_i$ associated with the latter channels are then rendered finite within the RG flow as a consequence of the fundamental quark-gluon interactions, i.e., via gluon-exchange box diagrams. This leads to further contributions in the flows for $\lambda_i$ proportional to $\lambda_j \lambda_l$ (fish diagrams) and $\lambda_j g^2$ (triangle diagrams), including also contributions from $\lambda_\textrm{top}$. As mentioned above, the couplings $\lambda_i$ associated with $U(1)_{\rm A}$-symmetric channels (or rather their flow) decay as $\sim 1/k^2$ for large cutoff scales. Their strength at lower scales is dominated by the integrated flow and their initial value should be set to zero. The four-quark couplings~${\lambda}_i$ are therefore not free parameters but exclusively generated by quark-gluon dynamics. 

The above arguments also entail that the contributions to the flow depending on the couplings $\lambda_i$ are subleading for perturbative momentum scales $k\gtrsim 3\,\text{GeV}$. For these scales, the flows for all $\lambda_i$, except for that of~$\lambda_\textrm{top}$, are dominated by the quark-gluon box diagrams $\sim\! g^4$. These contributions are largely driven by the growth of the strong coupling towards smaller scales~$k$ which stabilise the whole setup. For momentum scales $k\lesssim 2\,\text{GeV}$, however, the $\lambda_i$-contributions start to become dominant, see also our discussion in \autoref{sec:ua1T}.  

In the present work, we restrict ourselves to the pointlike limit of the four-quark correlation functions. The projection of the functions on the associated local four-quark couplings $\lambda_i$ in Eq.~\eqref{eq:delta4f} can then be defined as 
\begin{align}
\label{eq:def4f}
{\lambda}_{i}  = k^2\lim_{\{p_j\to 0\}}\Gamma^{(4)}_{(\bar{q}{\mathcal O}_iq)^2}(p_1,p_2,p_3,p_4)\,,
\end{align}
where the subscript $(\bar{q}{\mathcal O}_iq)^2$ stands for a suitable projection of the respective tensor channel, see, e.g., Refs.~\cite{Gies:2005as, Braun:2006jd,Braun:2011pp, Mitter:2014wpa, Cyrol:2017ewj, Braun:2018bik, Braun:2019aow}. Note that the commonly used Fierz-complete bases are in general not orthogonal.  

While we have removed the momentum-dependence of the channels with the projection~\eq{eq:def4f}, the scale dependence still carries an averaged momentum dependence. In the absence of strong angular dependences and rapid radial decays, this averaged momentum-dependence allows for even quantitative results, see, e.g., Refs.~\cite{Ellwanger:1994wy, Gies:2002hq, Gies:2005as, Braun:2005uj, Braun:2006jd, Braun:2011pp,Braun:2014wja, Braun:2014ata, Mitter:2014wpa, Rennecke:2015eba, Cyrol:2017ewj, Braun:2019aow, Fu:2019hdw}. Naturally, in regimes with emergent resonances the approximation~\eq{eq:def4f} gradually loses reliability and finally fails. For this reason, we restrict the current study to the chirally symmetric high-temperature regime of QCD. This is still sufficient for our present purposes as it already allows us to study the onset of spontaneous symmetry breaking, indicated by diverging four-quark couplings in the pointlike limit. Indeed, the latter observation also underlies many studies of the QCD phase structure with few and also many quark flavours~\cite{Gies:2005as, Braun:2005uj, Braun:2006jd, Braun:2011pp, Kusafuka:2011fd, Aoki:2014ola, Braun:2019aow}. For a detailed discussion of this aspect and the relation of pointlike four-quark couplings to the order-parameter potential of a given theory, we refer the reader to Refs.~\cite{Braun:2011pp, Braun:2017srn, Roscher:2019omd}.

We now focus again on two-flavour QCD. In this case, the standard scalar-pseudoscalar interaction channel associated with the formation of a chiral condensate,
\be
\left(\bar q q\right)^2\!-\! \left(\bar q \gamma_5 \tau_i q\right)^2\,,
\ee
is most dominant at low densities~\cite{Braun:2006zz,Mitter:2014wpa,Braun:2019aow}, signalling the spontaneous breakdown of the chiral symmetry in the low-energy limit. Here, $\left(\bar q \gamma_5 \tau_i q\right)^2\equiv \left(\bar q \gamma_5 \tau_i q\right)\left(\bar q \gamma_5 \tau_i q\right)$ and the~$\tau_i$'s are the Pauli matrices. Moreover, this channel explicitly breaks $U(1)_{\rm A}$ symmetry. This explicit breaking can be compensated by adding a four-quark channel as the one specified in Eq.~\eqref{eq:gua1}, 
\begin{align}
{\text{det}} \left[ \bar{q}_i P_{\rm L} q_j \right] + {\text{det}} \left[ \bar{q}_i P_{\rm R} q_j \right]\,,
\end{align}
provided that the value of the coupling associated with this channel is adjusted suitably relative to the scalar-pseudoscalar coupling~\cite{Braun:2011pp,Braun:2018bik,Braun:2019aow}. Thus, if there were only two four-quark channels, the relative strength of the corresponding couplings could be used to ``measure" the strength of explicit $U(1)_{\rm A}$ breaking. 

In practice, the situation is more involved. Even if the relative strength of these two couplings is not fine-tuned, the $U(1)_{\rm A}$ symmetry can still be restored by other four-quark interaction channels which are also invariant under $SU(\Nc)\otimes SU(\Nf)_\text{L}\otimes SU(\Nf)_\text{R}\otimes U(1)_\text{V}$ transformations. Therefore, an analysis of $U(1)_{\rm A}$ restoration in general requires to include all linearly-independent four-quark interactions compatible with the $SU(\Nc)\otimes SU(\Nf)_\text{L}\otimes SU(\Nf)_\text{R}\otimes U(1)_\text{V}$ symmetry. Taking also into account the explicit breaking of Poincar\'{e} invariance at finite temperature, the minimal Fierz-complete basis set of four-quark channels  for QCD with two {\it massless} quark flavours is composed of 10 channels in the pointlike limit~\cite{Braun:2018bik}, see \hyperref[app:2]{App.~\ref{app:2}}. For two {\it massless} flavours and one heavy flavour, we even end up with a Fierz-complete set of 32 four-quark interaction channels, see \hyperref[app:2+1]{App.~\ref{app:2+1}}. Taking also explicit chiral symmetry breaking in the light-quark sector into account, the number of possible channels increases even further.

In the following, we shall assume that an analysis of QCD with two massless quark flavours also provides useful information on effective $U(1)_{\rm A}$ restoration in the phenomenologically most relevant case of two light flavours and one heavy quark flavour. For our present studies, we therefore employ the two-flavour four-quark basis developed in Ref.~\cite{Braun:2018bik}. The corresponding basis elements can be found in \hyperref[app:2]{App.~\ref{app:2}}. The choice for these elements is motivated by channels conventionally used in phenomenological QCD models, such as the standard scalar-pseudoscalar channel associated with pion interactions, a diquark-type channel, and a channel of the form~\eqref{eq:gua1} associated with non-trivial topological gauge configurations. 

\subsection{Sum rules}\label{sec:sr_intro}
The 10-dimensional space of four-quark couplings in two-flavour QCD contains an 8-dimensional $U(1)_\text{A}$-invariant subspace, see \hyperref[app:2]{App.~\ref{app:2}} for the basis employed in our present work. While the first six tensors in \hyperref[app:2]{App.~\ref{app:2}}, see Eqs.~\eq{eq:ch1}-\eq{eq:ch6}, are manifestly $U(1)_\text{A}$-invariant, the other four admit two $U(1)_\text{A}$-invariant combinations. This gives rise to sum rules of the respective couplings~\cite{Braun:2018bik}:
{\allowdisplaybreaks
\begin{subequations}\label{eq:sumrules} 
\begin{eqnarray}
 {\lambda}_\Csc+ {\lambda}_\SpPmAdj &=& 0\,,\label{eq:sumrule1}\\
{\lambda}_{\text{top}} \!-\! \frac{\Nc \!-\! 1}{2\Nc}\,{\lambda}_\Csc \!+\!\frac{1}{2}{\lambda}_{(\sigma \text{-} \pi)} &=& 0\,.\label{eq:sumrule2}
\end{eqnarray}
\end{subequations}
Here, the} coupling~${\lambda}_{(\sigma \text{-} \pi)}$ is associated with the standard scalar-pseudoscalar four-quark channel \eq{eq:ch7}, the coupling~${\lambda}_{\text{top}}$ is associated with the ``topological channel" given in Eq.~\eqref{eq:gua1}, see also Eq.~\eq{eq:ch8}, and~${\lambda}_\Csc$ is associated with the conventional two-flavour diquark channel, see Eq.~\eq{eq:ch9}. Finally, the channel corresponding to the coupling~${\lambda}_\SpPmAdj$ may be viewed as the counterpart of the aforementioned ``topological channel" with a non-trivial colour structure, see Eq.~\eq{eq:ch10}. We emphasise that the $U(1)_{\text{A}}$ symmetry is only intact if both sum rules are satisfied simultaneously. 

The sum rules \eq{eq:sumrules} represent key ingredients for our study of $U(1)_{\text{A}}$ restoration at high temperature in \autoref{sec:ua1T} below. We add that such an analysis carries over straightforwardly to QCD with two massless flavours and one heavy flavour, but is beyond the scope of the present work. Here, we only remark that the associated 32-dimensional space of four-quark interactions contains a 26-dimensional $U(1)_{\rm A}$-invariant subspace, see \hyperref[app:2+1]{App.~\ref{app:2+1}}.

\subsection{Flow equations for the four-quark interactions}\label{sec:flow4quark}
Within our present approximations, the flow equations for the dimensionless renormalised four-quark couplings $\lambda_i$ assume the following form:
\begin{equation}
\partial_t\lambda_i =2\lambda_i - \sum_{jl}\lambda_j A^{(i)}_{jl} \lambda_l - \sum_{j} B^{(i)}_j \lambda_j g^2 - C^{(i)} g^4\,.
\label{eq:betafct}
\end{equation}
Here,~$t=\ln(k/\Lambda)$ and the indices~$i,j,l$ refer to elements of the Fierz-complete basis of four-quark couplings in \hyperref[app:2]{App.~\ref{app:2}}. The coefficients $A$ (fish diagram), $B$ (triangle diagram), and $C$ (box diagram) depend on the dimensionless temperature~$T/k$. 

We have dropped an implicit dependence of the loop diagrams on the wave function renormalisation factor of the gluon fields since it has been found to be subleading in the absence of spontaneous symmetry breaking~\cite{Gies:2002hq, Gies:2003dp, Gies:2005as, Braun:2005uj, Braun:2006jd}. Moreover, for convenience, we have restricted ourselves to the Feynman gauge. For our numerical analysis of the two-flavour case as presented in \autoref{sec:ua1T} below, we have derived the flow equations for the 10 four-quark couplings by using existing software packages~\cite{Huber:2011qr, Huber:2019dkb, Cyrol:2016zqb}, see also Ref.~\cite{Braun:2019aow} for details. Because of the size of the resulting system of differential equations, we only dealt with the flow equations numerically and therefore do not give an explicit representation of them here. However, for the limit of vanishing gauge coupling, the flow equations can be found in Ref.~\cite{Braun:2018bik}. For the case of an~$SU(\Nc)\otimes U(2)_{\rm L}\otimes U(2)_{\rm R}$ symmetry, an explicit representation of a Fierz-complete set of flow equations for the four-quark couplings in the vacuum limit is given in Ref.~\cite{Gies:2005as}, including the triangle and box diagrams. 

We add that we have used a covariant exponential regulator~\cite{Jungnickel:1995fp, Berges:1997eu, Berges:2000ew} in our numerical studies. For covariant regulators no spurious Lorentz-symmetry breaking is introduced via the regulator, and we are only left with the physical breaking due to the heat bath. In turn, for three-dimensional regulators unphysical Lorentz-symmetry breaking can occur, see, e.g., Refs.~\cite{Braun:2017srn, Pawlowski:2015mia, Pawlowski:2017gxj}. This issue is indeed of great relevance for Fierz-complete studies, see Ref.~\cite{Braun:2017srn} for a detailed discussion.  

The gauge sector enters our flow equations~\eqref{eq:betafct} for the four-quark interactions only via the running strong coupling~$g$. For the latter, we employ the results from Refs.~\cite{Braun:2005uj, Braun:2006jd}. There, the dependence of~$g$ on the RG scale~$k$ and the temperature~$T$ has been computed non-perturbatively within the fRG framework. Moreover, this has been done with the same exponential regulator as we use here in the flow equations for the four-quark couplings. Finally, we note that, in general, the running of the gauge coupling receives corrections from the four-quark couplings, see, e.g., Ref.~\cite{Gies:2003dp}. However, in the high-temperature regime, where the RG flow of the four-quark couplings is governed by the presence of fixed points, this back-reaction of the four-quark couplings on the RG flow of the gauge coupling has been found to be small, see, e.g., Refs.~\cite{Gies:2005as, Braun:2005uj, Braun:2006jd}. This completes the discussion of our setting. Subsequently, we now discuss chiral and effective $U(1)_{\rm A}$ restoration at finite temperature.

\section{Chiral and $U(1)_{\rm A}$ restoration at high temperatures}\label{sec:ua1T}
In this section, we present and discuss our results for chiral and $U(1)_{\rm A}$-symmetry breaking as well as their restorations at high temperatures. We start the analysis in \autoref{sec:GenResults} with a discussion of the general structure of the intertwined dynamics of four-quark interactions in QCD at the example of $U(1)_{\rm A}$-symmetric initial conditions. This already reveals the phenomenological mechanisms at work in this system.  The initial conditions, in particular those for the $U(1)_{\rm A}$-symmetry breaking coupling $\lambda_{\textrm{top}}$, are discussed and specified in \autoref{sec:InCondResults}. We then discuss $U(1)_{\rm A}$-symmetry breaking at zero temperature in~\autoref{sec:SumRulesResults} with the aid of the sum rules introduced in \autoref{sec:sr_intro} which provide us with a precise and phenomenological relevant definition of the strength of $U(1)_{\rm A}$-symmetry breaking. In particular, this allows us to discuss the symmetry breaking pattern and its infrared stability. Our results on chiral and effective $U(1)_{\rm A}$ symmetry restoration at finite temperature are finally presented in \autoref{sec:RestoreResults}.
\begin{figure}[t]
	\begin{center}
		\includegraphics[width=1\linewidth]{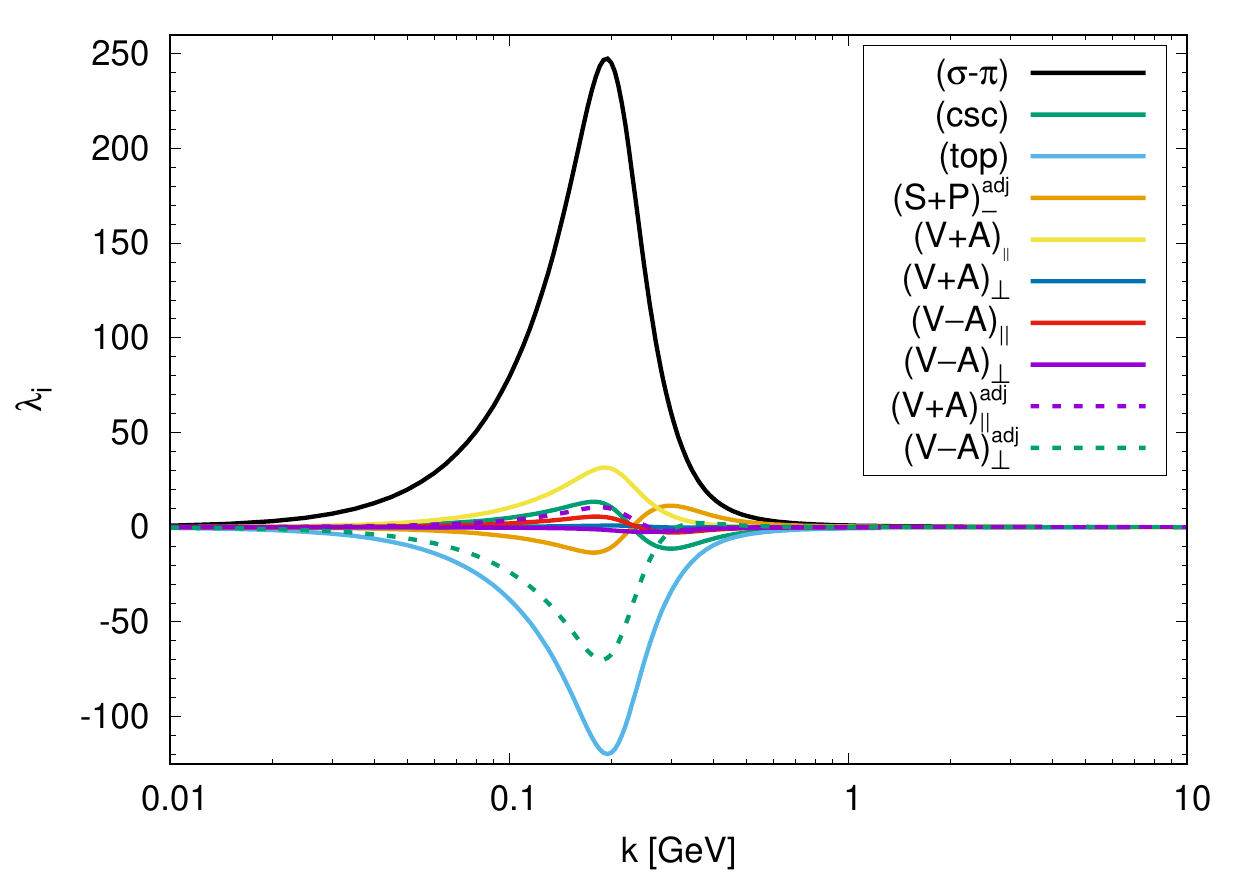}
	\end{center}
	\vspace*{-0.5cm}
	\caption{Scale dependence of the (dimensionless) four-quark couplings associated with the channels~\eqref{eq:ch1}-\eqref{eq:ch10} 
		as obtained slightly above the chiral phase transition in the $U(1)_{\rm A}$-symmetric limit.}
	\label{fig:cf}
\end{figure}
\subsection{General considerations of the RG flow of four-quark interactions}\label{sec:GenResults} 
We begin with a structural analysis of the dynamics in the $U(1)_{\rm A}$-symmetric limit. In this case, all four-quark couplings vanish at the initial cutoff scale $k=\Lambda$. Their high-energy dynamics is therefore only triggered by the quark-gluon box diagrams as represented by the last term on the right-hand side of Eq.~\eqref{eq:betafct}, $\lambda_i \sim g^4$. As discussed above, the $U(1)_{\rm A}$ symmetry is then preserved in the flow. Following the RG flow towards the low-energy regime, the strength of the four-quark interactions increases, hand in hand with the rise of the strong coupling. 

The RG flows of the four-quark couplings and the strong coupling are indeed intimately related. To be specific, if the strong coupling remains smaller than a critical value, then the four-quark self-interactions remain finite on all scales, as dictated by the fixed points of their flow equations~\eqref{eq:betafct}. In this case, QCD remains in the chirally symmetric regime on all scales. For example, this is the situation at sufficiently high temperatures. Note that both the critical value of the gauge coupling as well as the gauge coupling itself depend on the 
temperature~\cite{Braun:2005uj, Braun:2006jd}. On the other hand, if the gauge coupling exceeds a critical value, then the fixed points of the four-quark couplings are pushed into the complex plane~\cite{Gies:2003dp, Gies:2005as}, see Ref.~\cite{Braun:2011pp} for a review. In this case, the four-quark self-interactions become critical and the associated couplings start to grow rapidly in the RG flow, eventually approaching a divergence at a finite scale~$k=k_{\text{SB}}$. This scale is associated with the onset of spontaneous symmetry breaking. 

Indeed, as also discussed in the previous section, the observation of rapidly growing or even diverging four-quark couplings is an indicator for the formation of a condensate in the direction associated with the most dominant four-quark channel, see also Ref.~\cite{Roscher:2019omd} for a related discussion in the context of condensed-matter physics. This is the situation encountered at low temperatures. For example, in accordance with phenomenological expectations, we observe that the scalar-pseudoscalar channel is most dominant over a wide range of temperatures, indicating spontaneous chiral symmetry breaking at sufficiently low temperatures, see also \autoref{fig:cf} for an illustration. The critical temperature~$T_{\text{cr}}$ associated with chiral symmetry restoration is then defined as the smallest temperature for which the four-quark couplings remain finite on all scales~\cite{Braun:2005uj, Braun:2006jd, Braun:2011pp}. We add that, strictly speaking, this only defines an upper bound for the chiral phase transition temperature. Indeed, even at temperatures where we encounter a finite symmetry breaking scale~$k_{\text{SB}}$, the symmetries of the theory may still be restored in the IR limit by Goldstone fluctuations, see, e.g., Refs.~\cite{Braun:2009si,Braun:2011pp} for detailed discussions of this aspect.

\subsection{Initial conditions} \label{sec:InCondResults}

Let us now discuss the initial conditions which we employ in our analysis of effective $U(1)_{\rm A}$ restoration in QCD with two massless quark flavours below. As discussed in \autoref{sec:UA1}, nontrivial topological gauge field configurations induce $U(1)_{\rm A}$-violating four-quark interactions. This leads us to a non-vanishing 't Hooft coupling $\lambda_{\text{top}} (\Lambda)\neq 0$ at the initial cutoff scale $k=\Lambda$. In view of the study of symmetry restoration at finite temperature, we choose the initial scale $\Lambda$ large enough for having temperature-independent initial conditions for the temperature range of interest in this work, $T\lesssim 0.5\,\text{GeV}$. It has been shown in Refs.~\cite{Helmboldt:2014iya, Fister:2015eca, Braun:2018svj} that the thermal range of the exponential regulators considered here is $T_\textrm{max} /\Lambda\lesssim 5$. In other words, for temperatures $T\leq   T_\textrm{max} $, the initial conditions are temperature-independent. Moreover, such a choice additionally reduces the cutoff dependence, see Ref.~\cite{Braun:2018svj} for a general discussion of this aspect. 

With these considerations at hand, we can specify our initial conditions for the couplings in our present approximation, $(g,\{\lambda_i\})$, at the initial cutoff scale $\Lambda=10$\,GeV. Note that the part of the scale running of~$\lambda_{\text{top}} (k)$ triggered by fluctuations around topological configurations or lumps is genuinely non-perturbative and is of the form~$\sim {\rm e}^{-8\pi^2/g^2}$. This alone leads to a rapid rise of $\lambda_{\text{top}} (k)$ in the regime $k\sim 2 - 10$\,GeV, see \autoref{fig:lambdatop_top}.  As discussed in \autoref{sec:UA1}, we simply ``mimic" the respective scale-running in the present work by varying the initial 't~Hooft coupling $\lambda_{\text{top}} (\Lambda)$ in the range specified in Eq.~\eqref{eq:lambdatoprange}. Note that the reliability of this procedure has been already tested in Ref.~\cite{Hamada:2020mug} in a QCD-gravity setup. While the physics scales considered there are vastly different, the generic structure is the same. A full analysis including the topological running is deferred to future work.  

The initial values of the remaining nine four-quark couplings are set to zero. Finally, the initial condition for the gauge coupling is adjusted at zero temperature in the $U(1)_{\rm A}$-symmetric limit such that we obtain~$T_0=T_{\text{cr}}(\lambda_{\text{top}}(\Lambda)=0)\approx 0.132\,\text{GeV}$ for the chiral critical temperature. The so obtained value for the initial value for the strong coupling is then kept fixed when we vary~$\lambda_{\text{top}}(\Lambda)$. For our present purposes, this scale-fixing procedure is justified since $U(1)_{\rm A}$-violating effects are expected to be small in the perturbative high-energy regime where we fix the initial conditions. Moreover, this procedure ensures comparability between the theories associated with different values of~$\lambda_{\text{top}}(k=\Lambda)$. Our specific choice for the critical temperature~$T_0$ is motivated by recent lattice~\cite{Ding:2019prx} and fRG~\cite{Braun:2020ada} studies of QCD with two degenerate light quarks and physical strange quarks where~$T_{\text{cr}} \approx 0.132\dots 0.142\,\text{GeV}$ was found. In our numerical analysis below, we have employed the value from the lower end of this range as we expect the topologically effects associated with a variation of~$\lambda_{\text{top}}(\Lambda)$ to be most pronounced in this case.

\subsection{$U(1)_{\rm A}$-symmetry breaking and sum rules}\label{sec:SumRulesResults}
Within our present setting, we study the fate of~$U(1)_{\rm A}$ breaking by exploiting the sum rules~\eqref{eq:sumrules}. To be specific, we define
 \begin{subequations}\label{eq:SumRulesNorm}
 \begin{eqnarray} 
 S_1 &=& {\mathcal N}_1\left({\lambda}_\Csc+ {\lambda}_\SpPmAdj \right)\,,\label{eq:sumrule1norm}\\[1ex]
 S_2 &=& {\mathcal N}_2\left({\lambda}_{\text{top}} \!-\! \frac{\Nc \!-\! 1}{2\Nc}\,{\lambda}_\Csc \!+\!\frac{1}{2}{\lambda}_{(\sigma \text{-} \pi)} \right)\,,\label{eq:sumrule2norm}
\end{eqnarray} 
with appropriately chosen scale- and temperature-dependent normalisation factors, 
\begin{align}
{\mathcal N}_1=\frac{1}{2}{\lambda}_{(\sigma \text{-} \pi)}^{-1}\quad\text{and}\quad{\mathcal N}_2=\frac{2\Nc}{4\Nc-1}{\lambda}_{(\sigma \text{-} \pi)}^{-1}\,.
\label{eq:norm}
\end{align}
\end{subequations}
With the choice \eq{eq:norm} for~${\mathcal N}_1$ and~${\mathcal N}_2$, we simply measure the strength of $U(1)_{\rm A}$-violation relative to the coupling of the strongest four-quark channel which we found to be the scalar-pseudoscalar coupling. In our opinion, this is a phenomenologically very sensible measure of $U(1)_{\rm A}$-violation. In particular, it allows us to study (effective) $U(1)_{\rm A}$ restoration in a well-defined way. We will elaborate on this further below. Note also that we have $|S_i|\leq 1$ since the scalar-pseudoscalar coupling is found to be most dominant for all considered values of~$\lambda_{\text{top}}$.\footnote{We add that~$S_1$ and~$S_2$ are only bounded if the scalar-pseudoscalar coupling is non-zero which is the case at least towards the low-energy regime.}

From Eq.~\eqref{eq:sumrules}, we immediately deduce that the $U(1)_{\rm A}$ symmetry is only intact if we have~\mbox{$S_1=S_2=0$}. This is realised for $U(1)_{\rm A}$-symmetric boundary conditions where the~$U(1)_{\rm A}$ symmetry may only be broken dynamically below the symmetry breaking scale~$k_{\text{SB}}$. Any infinitesimally small violation of the $U(1)_{\rm A}$ symmetry at the initial scale immediately renders~$S_1$ and~$S_2$ finite. Moreover, this breaking is sustained in the RG flow for all scales. 
\begin{figure}[t]
	\begin{center}
		\includegraphics[width=1\linewidth]{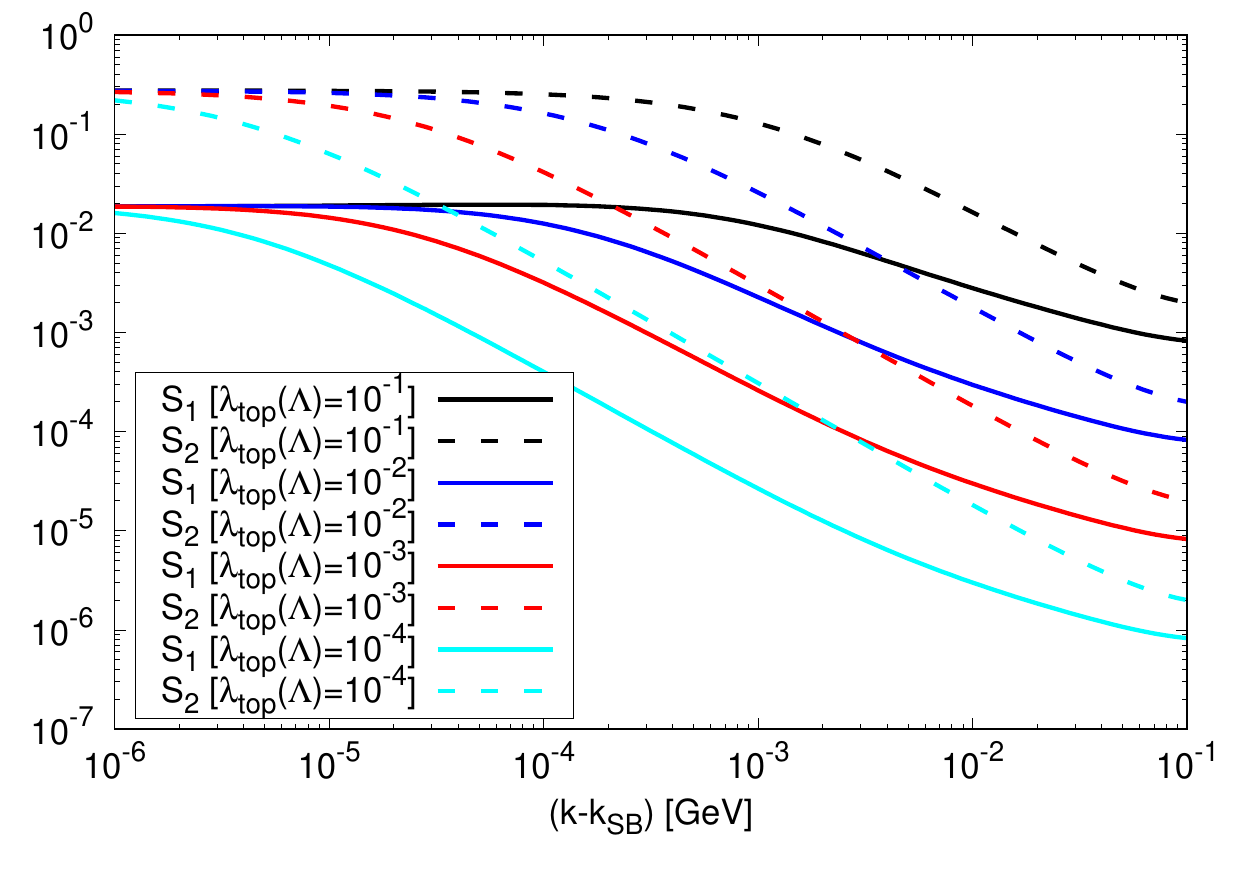}
	\end{center}
	\vspace*{-0.5cm}
	\caption{Scale dependence of the sum rules~\eqref{eq:sumrule1norm} and~\eqref{eq:sumrule2norm} at zero temperature for four different values of~$\lambda_{\text{top}}(\Lambda)$. Note that~$k_{\text{SB}}$ depends on~$\lambda_{\text{top}}(\Lambda)$. The $U(1)_{\rm A}$-symmetric limit corresponds to the case $S_1=S_2=0$ which is only realized for~$\lambda_{\text{top}}(\Lambda)=0$.}
	\label{fig:sr_T0}
\end{figure}

A strong deviation of~$S_1$ and~$S_2$ from zero indicates strong effective $U(1)_{\rm A}$ breaking. We emphasise, that in the presence of an explicit symmetry breaking, any investigation of effective $U(1)_{\rm A}$ restoration ultimately depends on the definition of the quantity used to ``measure"  the strength of the violation of the $U(1)_{\rm A}$ symmetry and is therefore not unique. Still, as we discuss below, the quantities~$S_1$ and~$S_2$  allow us to study the violation of the $U(1)_{\rm A}$ symmetry in a meaningful way above the chiral transition temperature and therefore provide us with an insight into the dynamics of QCD close to the chiral phase transition. 
\begin{figure*}[t]
	\begin{center}
		\includegraphics[width=0.5\linewidth]{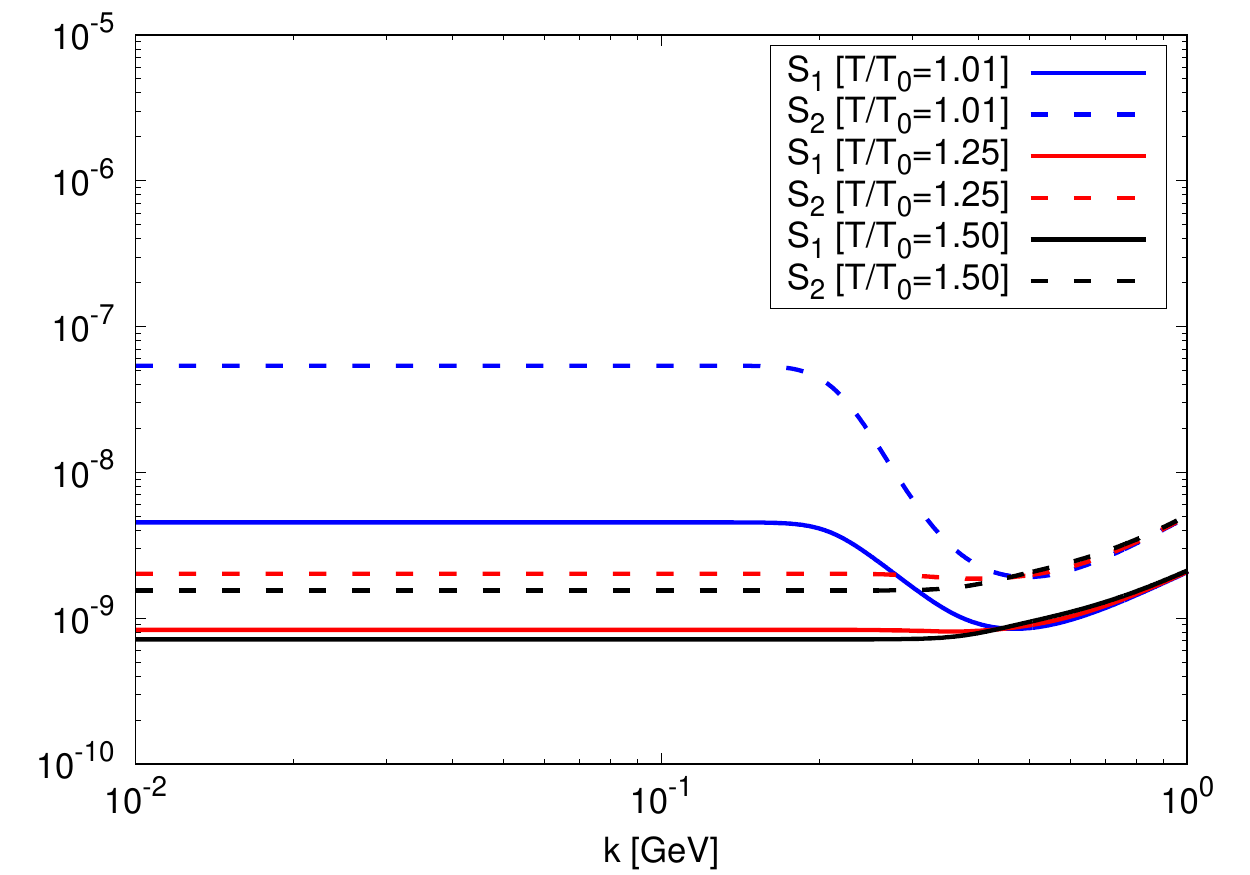}\hspace*{0.7cm}
		\includegraphics[width=0.5\linewidth]{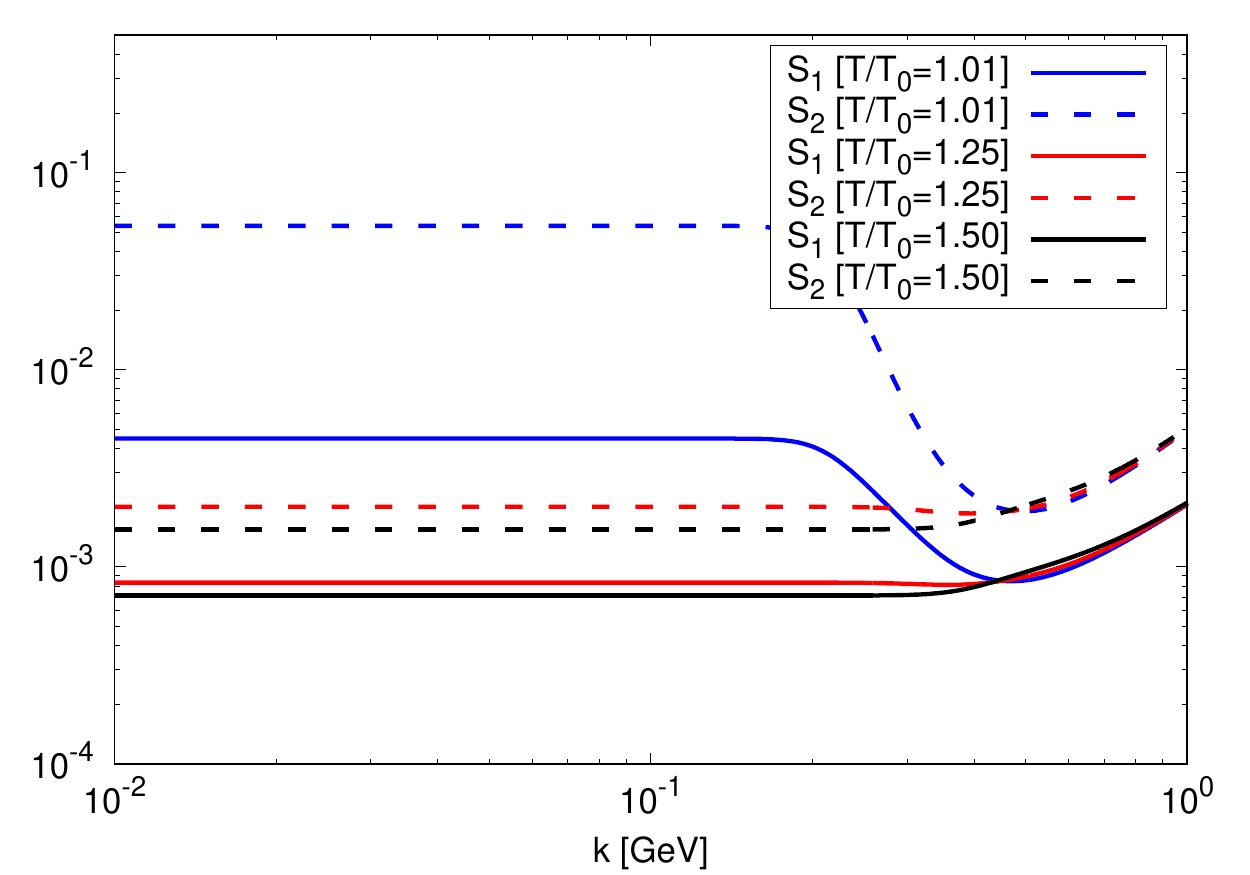}
	\end{center}
	\vspace*{-0.5cm}
	\caption{Scale dependence of the sum rules~\eqref{eq:sumrule1norm} and~\eqref{eq:sumrule2norm} for three different temperatures in the 
		chirally symmetric phase for~$\lambda_{\text{top}}(\Lambda)=10^{-7}$ (left panel) 
		and~$\lambda_{\text{top}}(\Lambda)=10^{-1}$ (right panel),  
		where~$T_0$ is the chiral phase transition temperature in the $U(1)_{\rm A}$-symmetric limit.}
	\label{fig:sr_finite_T}
\end{figure*}

Let us now discuss the symmetry-breaking pattern at vanishing temperature in the vacuum. Here, we find that the four-quark self-interactions become critical: the associated couplings start to grow rapidly and eventually diverge at a finite RG scale~$k=k_{\text{SB}}$. In the present setup we find~$k_{\text{SB}}\approx 0.346\,\text{GeV}$ in the $U(1)_{\rm A}$-symmetric limit. This is compatible with the symmetry-breaking scale in QCD within dynamical hadronisation, see Refs.~\cite{Braun:2014ata, Rennecke:2015eba, Mitter:2014wpa, Cyrol:2017ewj, Alkofer:2018guy, Fu:2019hdw}. In any case, the divergence at this scale signals the onset of spontaneous symmetry breaking. Moreover, as already mentioned above, we observe that the scalar-pseudoscalar coupling is most dominant for all considered values of~$\lambda_{\text{top}}(\Lambda)$, at least close to the symmetry breaking scale, for a quantitative analysis, see Refs.~\cite{Mitter:2014wpa, Cyrol:2017ewj}. In the latter studies, it has also been shown that the symmetry breaking scale within full QCD-flows is indeed very close to the divergence observed within the approximation with local four-quark interactions.

In \autoref{fig:sr_T0}, we show the scale dependence of the sum rules~$S_1$ and~$S_2$ for four different values of~$\lambda_{\text{top}}(\Lambda)$ at vanishing temperature. For example, for~$\lambda_{\text{top}}(\Lambda)=10^{-1}$ we have~$S_1\approx 0.019$ and~$S_2 \approx 0.25$ for~$k\to k_{\text{SB}}$, indicating a significant violation of the $U(1)_{\rm A}$ symmetry. Importantly, we also observe that both~$S_1$ and~$S_2$ become almost independent of the initial value of the 't~Hooft coupling~$\lambda_{\text{top}}(\Lambda)$ when the system approaches the symmetry breaking scale. In other words, the flow appears to lose its memory on the actual initial value of~$\lambda_{\text{top}}(\Lambda)$, at least for the range of values considered in this work. Thus, a finite value for $\lambda_{\text{top}}(\Lambda)$ is only required to initially break the~$U(1)_{\rm A}$ symmetry. This quasi fixed-point behaviour can be traced back to the existence of IR attractive fixed points of the four-quark couplings which govern the RG flow over a wide range of scales for sufficiently small values of the gauge coupling. Only if the gauge coupling becomes sufficiently large towards the low-energy regime, these fixed points are destabilised, eventually leading to spontaneous symmetry breaking, see also our discussion above and Refs.~\cite{Gies:2005as, Braun:2005uj, Braun:2006jd} for a corresponding discussion of the fixed-point structure of four-quark interactions. 

This behaviour is present for all considered input values of~$\lambda_{\text{top}}(\Lambda)$, including our most extreme choices. We emphasise that the considered range for $\lambda_{\text{top}}(\Lambda)$ effectively takes into account additional contributions from the topological flow. In conclusion, together with the analysis in Ref.~\cite{Hamada:2020mug}, the observed quasi fixed-point behaviour provides evidence that the robustness of the flow is also present for the full QCD dynamics which naturally includes the fluctuations around topological configurations. However, a respective analysis is deferred to future work. 

\subsection{Symmetry restoration at high temperatures}\label{sec:RestoreResults} 
Let us now study chiral and effective $U(1)_{\rm A}$ restoration at finite temperature. First of all, this requires a definition of the restoration temperature~$T_{\text{res}}$ associated with effective~$U(1)_{\rm A}$ restoration. For this purpose, we use the normalised sum rules \eq{eq:SumRulesNorm} and define the $U(1)_{\rm A}$ symmetry to be restored if the combined \textit{normalised} breaking of both sum rules is below a certain threshold~$S_{\text{cr}}$. To be specific, we define $T_\textrm{res}$ as follows:
\begin{subequations}\label{eq:Tres}
\begin{align}\label{eq:DefTres}
T>T_{\text{res}}: \quad S_{\Sigma}(T) \leq S_{\text{cr}}\,,
\end{align} 
with the combined \textit{normalised} symmetry breaking~$S_{\Sigma}$,  
\begin{align}\label{eq:Srestore}
S_{\Sigma}=\frac12\left( |S_1|+|S_2|\right) \,. 
\end{align}
\end{subequations}
For temperatures~$T>T_{\text{res}}$, we then consider the $U(1)_{\rm A}$ symmetry to be effectively restored. In the present work, we consider a $0.5\%$-threshold as a sensible measure, i.e., $S_{\text{cr}}=0.005$. It entails that the combined sum rules are broken by less than one percent in terms of the most dominant channel.

We emphasise again that, for an explicitly broken symmetry, it is not possible to provide a unique definition of (effective) $U(1)_{\rm A}$ restoration. However, our present criterion relies on the analysis of the strength of~$U(1)_{\rm A}$-symmetry breaking relative to the most dominant four-quark  channel. In this spirit, our criterion based on the value of~$S_{\Sigma}$ relative to~$S_{\text{cr}}$ is a phenomenologically sound criterion for effective $U(1)_{\rm A}$ restoration.

\begin{figure}[t]
	\begin{center}
		\includegraphics[width=1\linewidth]{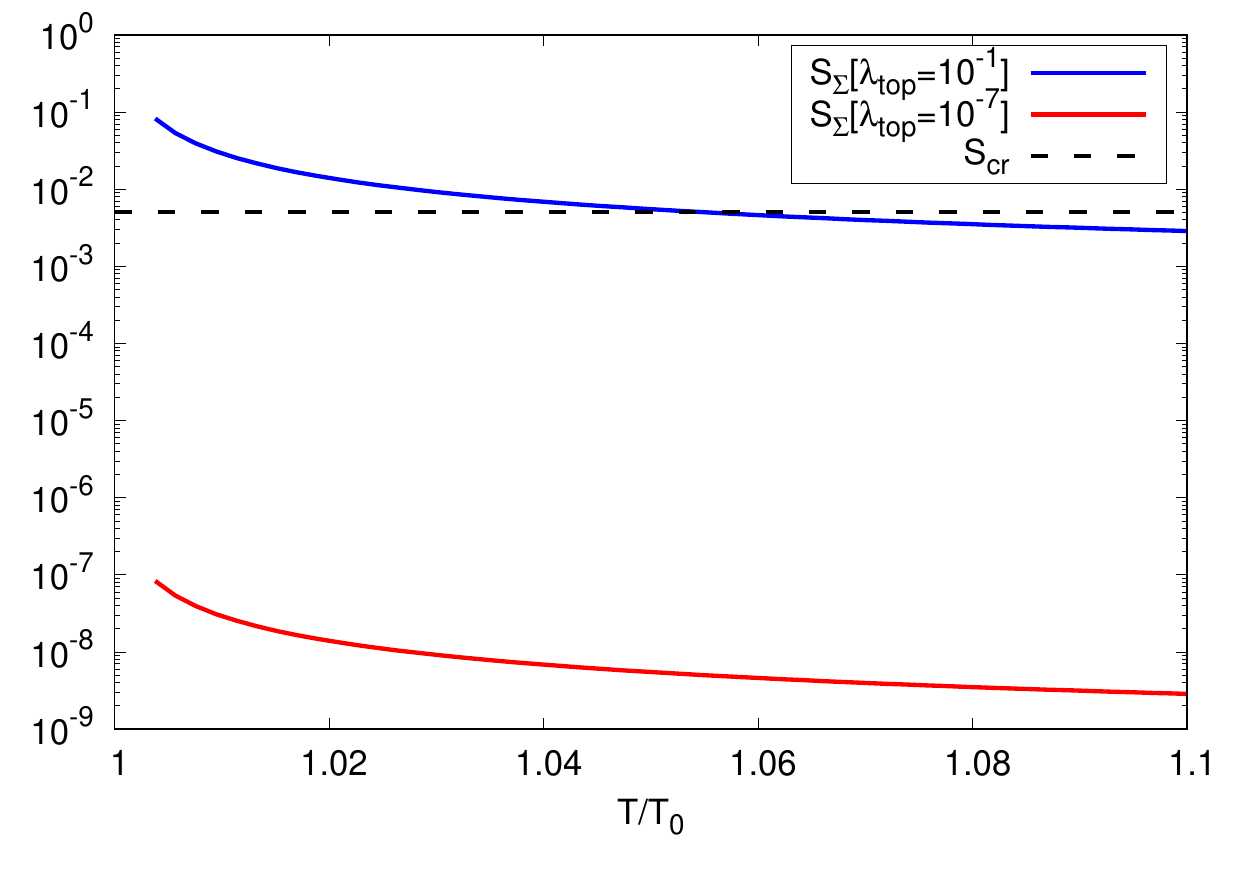}
	\end{center}
	\vspace*{-0.5cm}
	\caption{Temperature dependence of the IR value of the symmetry breaking parameter~$S_{\Sigma}$ for two different values of the initial 't Hooft coupling~$\lambda_{\text{top}}(\Lambda)$. Note that~$T_0$ is the chiral phase transition temperature in the $U(1)_{\rm A}$-symmetric limit.} 
	\label{fig:sr_T_dep}
\end{figure}

In \autoref{fig:sr_finite_T}, the temperature dependence of~$S_1$ and~$S_2$ as a function of the RG scale~$k$ is illustrated for~$\lambda_{\text{top}}(\Lambda)=10^{-7}$ (left panel) and~$\lambda_{\text{top}}(\Lambda)=10^{-1}$ (right panel) in the chirally symmetric high-temperature regime. For both initial values of~$\lambda_{\text{top}}(\Lambda)$, we observe that the IR values of~$S_1$ and~$S_2$ decrease monotonically with increasing temperature, indicating an effective restoration of the $U(1)_{\rm A}$ symmetry at high temperatures. This can also be seen in \autoref{fig:sr_T_dep}. There,  
the temperature dependence of the IR value of the symmetry breaking parameter~$S_{\Sigma}$ is presented for those two values of the 't Hooft coupling also shown in \autoref{fig:sr_finite_T}. 

The observed effective $U(1)_{\rm A}$ restoration at high temperatures can be traced back to the fact that quark fluctuations become more and more thermally suppressed with increasing temperature, eventually resulting in a decoupling of the gauge and matter sector. However, although the strength of $U(1)_{\rm A}$ breaking is almost identical for both values of~$\lambda_{\text{top}}(\Lambda)$ close to the chiral symmetry breaking scale in the zero-temperature limit (see our discussion of the quasi fixed-point behaviour above), we find that it requires higher temperatures to effectively restore the $U(1)_{\rm A}$ symmetry when~$\lambda_{\text{top}}(\Lambda)$ is increased. To better understand this observation, we consider again our results for~$S_1$ and~$S_2$ in the zero-temperature limit. Since the thermal Matsubara masses of the quarks act as an IR cutoff, the behaviour of a system at finite temperature may indeed be anticipated from its RG flow at zero temperature evaluated at~$k\sim T$. Looking at the zero-temperature flows depicted in \autoref{fig:sr_T0}, we notice that the dynamical enhancement of the initial $U(1)_{\rm A}$ breaking already sets in at higher scales~$k$ when we increase~$\lambda_{\text{top}}(\Lambda)$. Taking this into account, we conclude that our finite-temperature results reflect the scale dependence observed in the zero-temperature limit.

\begin{figure}[t]
	\begin{center}
		\includegraphics[width=1\linewidth]{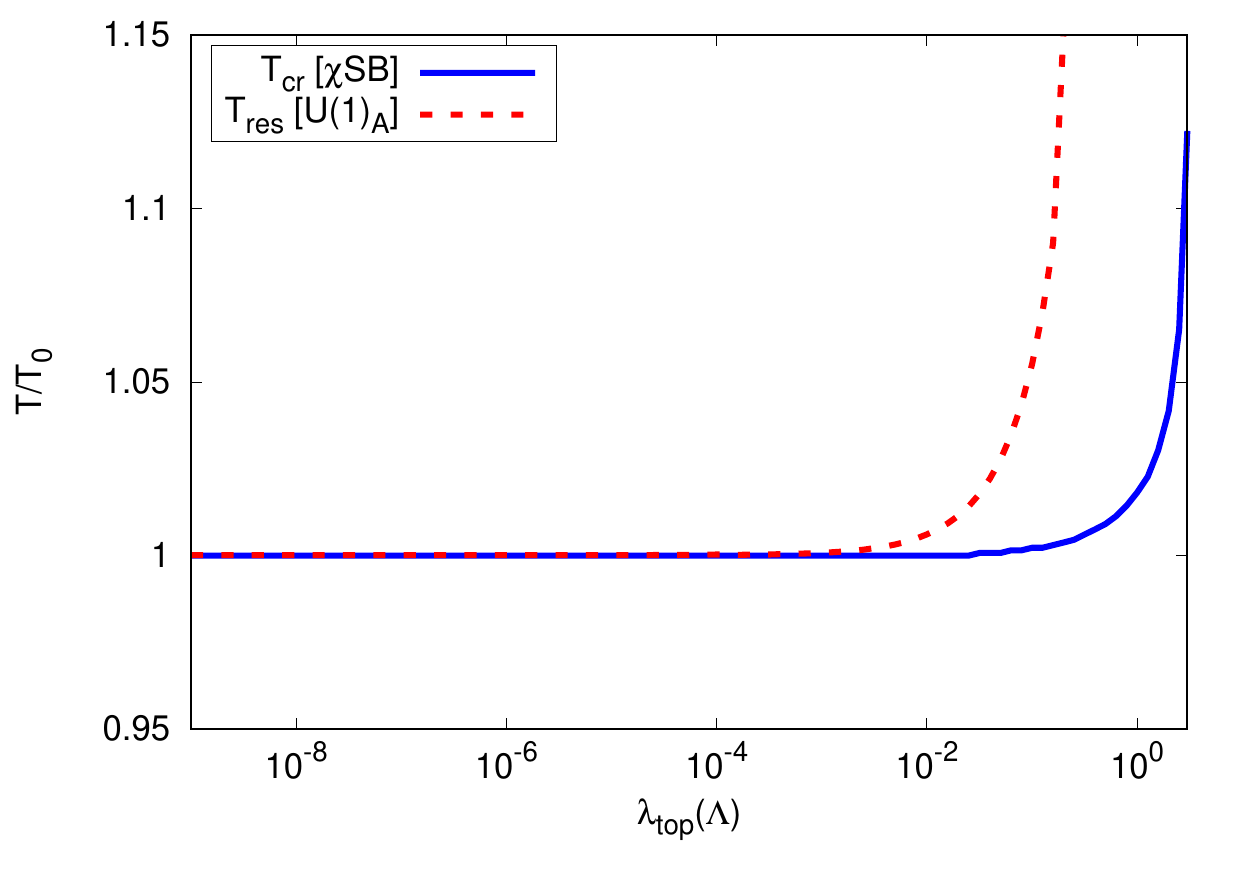}
	\end{center}
	\vspace*{-0.5cm}
	\caption{Estimates for the chiral phase transition temperature $T_{\text{cr}}\equiv T_{\text{cr}}[\chi\text{SB}]$ and the temperature~$T_{\text{res}}\equiv T_{\text{res}}[U(1)_{\rm A}]$ associated with effective $U(1)_{\rm A}$ restoration 
		as a function of the initial 't Hooft coupling~$\lambda_{\text{top}}(\Lambda)$. Here,~$T_0$ is again the chiral transition temperature in the $U(1)_{\rm A}$-symmetric limit.}
	\label{fig:pd}
\end{figure}

The dynamics at high temperatures in the chirally symmetric regime can be quantified by computing the temperature~$T_{\text{res}}$ associated with effective $U(1)_{\rm A}$ restoration, see Eq.~\eqref{eq:Tres} for its definition. In \autoref{fig:pd}, we show our results for~$T_{\text{res}}$ together with those for the chiral phase transition temperature~$T_{\text{cr}}$ as a function of~$\lambda_{\text{top}}(\Lambda)$. We observe that these two temperatures agree in the $U(1)_{\rm A}$-symmetric limit and are still very close to each other over many orders of magnitude of the 't Hooft coupling~$\lambda_{\text{top}}(\Lambda)$. For~$\lambda_{\text{top}}(\Lambda)\gtrsim 10^{-3}$, the chiral phase transition temperature~$T_{\text{cr}}$ and the $U(1)_{\rm A}$-restoration temperature then start to deviate, opening up a window of temperatures within the chiral symmetry is already restored but the $U(1)_{\rm A}$ symmetry is still significantly broken. 

The observation of an almost coincidence of chiral and effective $U(1)_{\rm A}$ symmetry restoration over a wide range of values of the initial 't Hooft coupling is of great interest from a phenomenological standpoint. In fact,  it has been shown that $U(1)_{\rm A}$ restoration at the phase transition 
affects directly the order and the universality class of the chiral transition~\cite{Pisarski:1983ms, Mitter:2013fxa, Grahl:2013pba,Pelissetto:2013hqa}. More specifically, a second-order phase transition in the~$O(4)$-universality class is only expected if the chiral transition and the effective restoration of the $U(1)_{\rm A}$ symmetry are sufficiently separated in temperature which we find to be the case for~$\lambda_{\text{top}}(\Lambda)\gtrsim 10^{-1}$. Taking into account results from explicit computations of the topological contributions to the running of the 't~Hooft coupling~\cite{Pawlowski:1996ch}, we therefore presently consider the~$O(4)$ scenario to be more likely as the topological running neglected in our present study tends to increase~$\lambda_{\text{top}}$ at high scales, see also \autoref{fig:lambdatop_top}. However, we rush to add that our present analysis does not allow us to make conclusive statements in this respect but rather calls for more detailed studies, including a study of the quark mass dependence of our present observations. In any case, with respect to QCD low-energy models, our study already reveals that the inclusion of four-quark channels other than the scalar-pseudoscalar channel and the 't~Hooft channel is relevant to improve their predictive power with respect to studies of the nature of the chiral transition. In fact, we find that $U(1)_{\rm A}$-breaking four-quark channels other than the aforementioned two channels are dynamically generated in the RG flow and carry relevant information on the fate of the~$U(1)_{\rm A}$ symmetry even above the chiral phase transition, see also \autoref{fig:cf}. Note that this effect is only partially due to quark-gluon interactions. In fact, choosing a single four-quark coupling, such as the one of the scalar-pseudoscalar channel, to be ``overcritical" immediately leads to a dynamical generation of all other four-quark channels compatible with the symmetries of the theory, see also Ref.~\cite{Braun:2018bik}. 
\section{Conclusions}\label{sec:conc}

In this paper we have presented an analysis of chiral and effective $U(1)_{\rm A}$ restoration at finite temperature in QCD with two massless quark flavours. For the study of these genuinely non-perturbative and intertwined phenomena we have used the fRG approach. More specifically, we have studied the RG flow of a Fierz-complete set of four-quark interactions starting from the classical QCD action in the high-energy limit. Topological fluctuations are well-known to generate $U(1)_{\rm A}$-violating four-quark interactions~\cite{tHooft:1976rip, tHooft:1976snw, THOOFT1986357}. To effectively include the effect of such fluctuations, we have varied the initial value of the corresponding four-quark coupling over a wide range of values in our RG analysis, including the physically relevant range. The latter has been estimated with the results of an earlier study of the purely topological running of this coupling~\cite{Pawlowski:1996ch}. 

In accordance with previous studies~\cite{Braun:2006zz, Mitter:2014wpa, Cyrol:2017ewj, Braun:2019aow}, our study reveals the dominance of the scalar-pseudoscalar interaction channel. The divergence of the corresponding coupling signals the onset of spontaneous chiral symmetry breaking in the low-energy regime and has been shown to be very close to the symmetry breaking scale in full QCD, see, in particular, Ref.~\cite{Mitter:2014wpa}. Since the strong coupling and the coupling of the topological four-quark channel are the only input parameters of our present analysis, this dominance is a nontrivial observation which cannot be traced back to a specific choice for the initial conditions. It is rather generated dynamically by the underlying quark-gluon dynamics. In fact, this dominance is even a robust feature for the considered range of values of the topological coupling. 

It is well-known that the analysis of $U(1)_{\rm A}$ restoration requires some care. In fact, since the $U(1)_{\rm A}$ symmetry is explicitly broken, any study of $U(1)_{\rm A}$ restoration depends on the definition of the quantity used to ``measure"  the strength of $U(1)_{\rm A}$ breaking. Consequently, in the presence of an explicit symmetry breaking the notion of symmetry restoration is not unique. For our present analysis of effective $U(1)_{\rm A}$ restoration, we exploited the fact that our Fierz-complete set of four-quark interactions contains a $U(1)_{\rm A}$-invariant subspace which can be defined via sum rules for the four-quark interactions. If these sum rules are simultaneously fulfilled, then the $U(1)_{\rm A}$ symmetry is intact. In turn, violations of these sum rules, measured in terms of the interaction strength of the most dominant channel, can be used to quantify the strength of $U(1)_{\rm A}$ breaking and estimate a temperature associated with effective $U(1)_{\rm A}$ restoration. Importantly, employing these rum rules, our analysis suggests a quasi fixed-point behaviour at zero temperature: close to the chiral symmetry breaking scale, we find that the violations of these rules appear almost independent of the considered values of the topological coupling. This observation can be traced back to the fixed-point structure of the four-quark couplings which is generated by quark-gluon interactions and causes a ``memory loss" with respect to the initial conditions. Note that the observed insensitivity with respect to a variation of the 't Hooft coupling is in accordance with an early detailed QCD model study~\cite{Jungnickel:1995fp}.

At finite temperature, we find that the $U(1)_{\rm A}$ symmetry is effectively restored at the chiral phase transition temperature over a wide range of initial values of the 't Hooft coupling, see \autoref{fig:pd}. For larger initial values of this coupling, $\lambda_\textrm{top}\gtrsim 10^{-2}$, we then observe that effective $U(1)_{\rm A}$ restoration and the restoration of the chiral symmetry are increasingly separated. Note that the location of the effective $U(1)_{\rm A}$ restoration directly affects the order and the universality class of the chiral transition~\cite{Pisarski:1983ms, Mitter:2013fxa, Grahl:2013pba, Pelissetto:2013hqa}. In particular, only for a clear separation of the chiral phase transition temperature and the restoration temperature $T_\textrm{res}$, we expect QCD to fall into the $O(4)$ universality class. Since the results for the purely topological running of the 't Hooft coupling~\cite{Pawlowski:1996ch} indeed indicate an \textit{effective} initial coupling of  $\lambda_\textrm{top}\gtrsim 10^{-2}$, our present analysis suggests that the $O(4)$ scenario is favoured. 

At present, of course, this is only an indication for the realisation of the $O(4)$ scenario at the phenomenologically relevant point in our phase diagram spanned by the temperature and the 't Hooft coupling. Our present work cannot conclusively resolve the issue regarding the order and the universality class of the QCD phase transition in the zero-density limit. However, the significant advances made within functional QCD approaches in recent years together with the present Fierz-complete study lay the ground for future quantitative fRG studies of the nature of the chiral phase transition in two-flavour QCD.

{\it Acknowledgments.--~} The authors as members of the fQCD collaboration~\cite{fQCD} thank the other members of this collaboration for discussions and work on related subjects. We also thank W.-j.~Fu, F.~Rennecke and B.-J. Schaefer for discussions. JB acknowledges support by the Deutsche Forschungsgemeinschaft (DFG, German Research Foundation) under grant BR 4005/4-1 (Heisenberg program) and by the Helmholtz International Center for the Facility for Antiproton and Ion Research (HIC for FAIR) within the LOEWE program of the State of Hesse. JB and DR acknowledge support by the DFG -- Projektnummer 279384907 -- SFB/TR 211. JB and ML acknowledge support by the DFG -- Projektnummer 279384907 -- SFB 1245. JMP is supported by EMMI, the BMBF grant 05P18VHFCA, and by the DFG Collaborative Research Centre SFB 1225 (ISOQUANT). This work is supported by Deutsche Forschungsgemeinschaft (DFG, German Research Foundation) under Germany's Excellence Strategy EXC-2181/1 - 390900948 (the Heidelberg STRUCTURES Cluster of Excellence).
\appendix
\section{Four-quark basis for $2$-flavour QCD}\label{app:2}

Here, we present the Fierz-complete basis of four-quark interaction channels employed in our present study of QCD with two massless flavours. This basis is 10-dimensional and 
can be written with the aid of the following basis elements~\cite{Braun:2018bik}:
{\allowdisplaybreaks
	\begin{eqnarray}
		\mathcal{L}_\VpAPar&=&\left(\bar q\gamma_0q\right)^2+\left(\bar q\I\gamma_0\gamma_5q\right)^2\,,\label{eq:ch1}
		\\[1ex]
		\mathcal{L}_\VpAPer&=&\left(\bar q\gamma_iq\right)^2+\left(\bar q\I\gamma_i\gamma_5q\right)^2\,,\label{eq:ch2}\\[1ex]
		\mathcal{L}_\VmAPar&=&\left(\bar q\gamma_0q\right)^2-\left(\bar q\I\gamma_0\gamma_5q\right)^2\,,\label{eq:ch3}\\[1ex]
		\mathcal{L}_\VmAPer&=&\left(\bar q\gamma_iq\right)^2-\left(\bar q\I\gamma_i\gamma_5q\right)^2\,,\label{eq:ch4}\\[1ex]
		\mathcal{L}_\VpAParAdj&=&\left(\bar q\gamma_0 T^aq\right)^2+\left(\bar q\I\gamma_0\gamma_5 T^aq\right)^2\,,\label{eq:ch5}\\[1ex]
		\mathcal{L}_\VmAPerAdj&=&\left(\bar q\gamma_iT^aq\right)^2-\left(\bar q\I\gamma_i\gamma_5 T^aq\right)^2\,,\label{eq:ch6}\\[1ex]
		\mathcal{L}_{\text{($\sigma $-$\pi $)}}&=&\left(\bar q q\right)^2\!-\! \left(\bar q \gamma_5 \tau_i q\right)^2\,,\label{eq:ch7}\\[1ex]
		\mathcal{L}_{\text{top}}&=& {\text{det}} \left[ \bar{q}_i P_{\rm L} q_j \right] + {\text{det}} \left[ \bar{q}_i P_{\rm R} q_j \right]\,,\label{eq:ch8}\\[1ex]
		\mathcal{L}_\Csc &=& 4 \left( \I \bar q \gamma_5 \tau_2\, T^{A} q^C \right) \left( \I \bar q^C \gamma_5 \tau_2\, T^{A} q \right)\,,\label{eq:ch9}\\[1ex]
		\mathcal{L}_{(S+P)_{-}^\mathrm{adj}} &=&\left(\bar q T^aq\right)^2\!-\!\left(\bar q \gamma_5 \tau_i T^aq\right)^2\nn \\
		&&\hspace{0.5 cm}\!+\!\left(\bar q \gamma_5 T^aq\right)^2\!-\!\left(\bar q \tau_i T^aq\right)^2\,,\label{eq:ch10}
\end{eqnarray}
where} the $T^a$'s are the generators of $SU(\Nc)$. Note that this choice for the basis elements~${\mathcal L}_i$ is not unique. It is rather motivated by channels conventionally used in phenomenological QCD models~\cite{Braun:2018bik}, see also our discussion in the main text. Note that the first six channels of this basis have been chosen to be invariant under  $SU(\Nc)\otimes SU(2)_\text{L}\otimes SU(2)_\text{R}\otimes U(1)_\text{V} \otimes U(1)_\text{A}$ transformations, whereas  the remaining four channels in Eqs.~\eqref{eq:ch7}-\eqref{eq:ch10}  explicitly break the $U(1)_{\rm A}$ symmetry. 

\section{Four-quark basis for $(2\!+\!1)$-flavour QCD}\label{app:2+1}
Here, we present a Fierz-complete basis of four-quark interaction channels for QCD with two massless and one heavy quark flavour. The $U(1)_{\rm A}$-symmetric limit of this system is spanned by a minimal set of 26 four-quark interaction channels which may be divided into interactions in the subspace of the two massless quark flavours,
{\allowdisplaybreaks
\be\nonumber 
\mathcal{L}_{(V+A)_{\parallel, l}} &=& \left( \bar{q} \gamma_0  \mathbbm{1}^{l} q \right)^2 + \left( \bar{q} \I \gamma_0 \gamma_5  \mathbbm{1}^{l} q \right)^2\,,  \\[1ex]\nonumber 
\mathcal{L}_{(V+A)_{\bot, l}} &=& \left( \bar{q} \gamma_i  \mathbbm{1}^{l} q \right)^2 + \left( \bar{q} \I \gamma_i \gamma_5  \mathbbm{1}^{l} q \right)^2\,,   \\[1ex]\nonumber 
\mathcal{L}_{(V-A)_{\parallel, l}} &=& \left( \bar{q} \gamma_0  \mathbbm{1}^{l} q \right)^2 - \left( \bar{q} \I \gamma_0 \gamma_5  \mathbbm{1}^{l} q \right)^2\,,  \\[1ex]\nonumber 
\mathcal{L}_{(V-A)_{\bot, l}} &=& \left( \bar{q} \gamma_i  \mathbbm{1}^{l} q \right)^2 - \left( \bar{q} \I \gamma_i \gamma_5  \mathbbm{1}^{l} q \right)^2\,,  \\[1ex]\nonumber 
\mathcal{L}_{(V+A)^{\text{adj}}_{\parallel, l}} &=& \left( \bar{q} \gamma_0 T^a  \mathbbm{1}^{l} q \right)^2 + \left( \bar{q} \I \gamma_0 \gamma_5 T^a  \mathbbm{1}^{l} q \right)^2\,,  \\[1ex]\nonumber 
\mathcal{L}_{(V+A)^{\text{adj}}_{\bot, l}} &=& \left( \bar{q} \gamma_i T^a \mathbbm{1}^{l} q \right)^2 + \left( \bar{q} \I \gamma_i \gamma_5 T^a  \mathbbm{1}^{l} q \right)^2\,, \\[1ex]\nonumber 
\mathcal{L}_{(V-A)^{\text{adj}}_{\parallel, l}} &=& \left( \bar{q} \gamma_0 T^a  \mathbbm{1}^{l} q \right)^2 - \left( \bar{q} \I \gamma_0 \gamma_5 T^a  \mathbbm{1}^{l} q \right)^2\,,  \\[1ex]
\mathcal{L}_{(V-A)^{\text{adj}}_{\bot, l}} &=& \left( \bar{q} \gamma_i T^a \mathbbm{1}^{l} q \right)^2 - \left( \bar{q} \I \gamma_i \gamma_5 T^a  \mathbbm{1}^{l} q \right)^2\,, 
\ee
and} interactions in the strange-quark subspace,
{\allowdisplaybreaks
\be\nonumber 
\mathcal{L}_{(V+A)_{\parallel, s}} &= &\left( \bar{q} \gamma_0  \mathbbm{1}^{s} q \right)^2 + \left( \bar{q} \I \gamma_0 \gamma_5  \mathbbm{1}^{s} q \right)^2\,,  \\[1ex]\nonumber 
\mathcal{L}_{(V-A)_{\parallel, s}} &=& \left( \bar{q} \gamma_0  \mathbbm{1}^{s} q \right)^2 - \left( \bar{q} \I \gamma_0 \gamma_5  \mathbbm{1}^{s} q \right)^2\,,  \\[1ex]\nonumber 
\mathcal{L}_{(V+A)^{\text{adj}}_{\bot, s}} &=& \left( \bar{q} \gamma_i T^a \mathbbm{1}^{s} q \right)^2 + \left( \bar{q} \I \gamma_i \gamma_5 T^a  \mathbbm{1}^{s} q \right)^2\,, \\[1ex]\nonumber 
\mathcal{L}_{(V-A)^{\text{adj}}_{\bot, s}} &=& \left( \bar{q} \gamma_i T^a  \mathbbm{1}^{s} q \right)^2 - \left( \bar{q} \I \gamma_i \gamma_5 T^a  \mathbbm{1}^{s} q \right)^2, \\[1ex]\nonumber 
\mathcal{L}_{(S-P)_{s}} &=& \left( \bar{q}  \mathbbm{1}^{s} q \right)^2 - \left( \bar{q} \gamma_5  \mathbbm{1}^{s} q \right)^2 \,, \\
\mathcal{L}_{(S-P)^{\text{adj}}_{s}} &=& \left( \bar{q}  T^a \mathbbm{1}^{s} q \right)^2 - \left( \bar{q}  \gamma_5 T^a \mathbbm{1}^{s} q \right)^2\,, 
\ee
and} interactions ``living" in the whole three-flavour space,
{\allowdisplaybreaks
\be\nonumber 
\mathcal{L}_{(V+A)_\parallel} &=& \left( \bar{q} \gamma_0 q \right)^2 + \left( \bar{q} \I \gamma_0 \gamma_5 q \right)^2\,,  \\[1ex]\nonumber 
\mathcal{L}_{(V+A)_\bot} &=& \left( \bar{q} \gamma_i q \right)^2 + \left( \bar{q} \I \gamma_i \gamma_5 q \right)^2\,,  \\[1ex]\nonumber 
\mathcal{L}_{(V-A)_\parallel} &=& \left( \bar{q} \gamma_0 q \right)^2 - \left( \bar{q} \I \gamma_0 \gamma_5 q \right)^2\,,  \\[1ex]\nonumber 
\mathcal{L}_{(V-A)_\bot} &=& \left( \bar{q} \gamma_i q \right)^2 - \left( \bar{q} \I \gamma_i \gamma_5 q \right)^2\,,  \\[1ex]\nonumber 
\mathcal{L}_{(V+A)^{\text{adj}}_\parallel} &=& \left( \bar{q} \gamma_0 T^a q \right)^2 + \left( \bar{q} \I \gamma_0 \gamma_5 T^a q \right)^2\,,  \\[1ex]\nonumber 
\mathcal{L}_{(V+A)^{\text{adj}}_\bot} &=& \left( \bar{q} \gamma_i T^aq \right)^2 + \left( \bar{q} \I \gamma_i \gamma_5 T^a q \right)^2\,, \\[1ex]\nonumber 
\mathcal{L}_{(V-A)^{\text{adj}}_\parallel} &=& \left( \bar{q} \gamma_0 T^a q \right)^2 - \left( \bar{q} \I \gamma_0 \gamma_5 T^a q \right)^2\,, \\[1ex]\nonumber 
\mathcal{L}_{(V-A)^{\text{adj}}_\bot} &=& \left( \bar{q} \gamma_i T^a q \right)^2 - \left( \bar{q} \I \gamma_i \gamma_5 T^a q \right)^2\,, \\[1ex]\nonumber 
\mathcal{L}_{(S-P)_{\mathbbm{1} + \textrm{ISO}}} &=& \left( \bar{q} q \right)^2 - \left( \bar{q} \gamma_5 q \right)^2 \nn \\[1ex]\nonumber 
&&\;\; + \sum_{k=1}^3 \left[  \left( \bar{q} \tau_k q \right)^2 - \left( \bar{q} \gamma_5 \tau_k q \right)^2 \right] \,, \\[1ex]\nonumber 
\mathcal{L}_{(S-P)^{\text{adj}}_{\mathbbm{1} + \textrm{ISO}}} &=& \left( \bar{q} T^a q \right)^2 - \left( \bar{q} \gamma_5 T^a q \right)^2 \nn  \\[1ex]\nonumber 
&& \!\! + \sum_{k=1}^3 \left[  \left( \bar{q} \tau_k T^aq \right)^2 - \left( \bar{q} \gamma_5 \tau_k T^a q \right)^2 \right] \,,  \\\nonumber 
\mathcal{L}_{T_{\bot - \parallel, \mathbbm{1} + \textrm{ISO}}} &=& \left( \bar{q} \sigma_{ij} q \right)^2 - 2 \left( \bar{q} \sigma_{0i}  q \right)^2 \nn  \\[1ex]\nonumber 
&& \!\! + \sum_{k=1}^3 \left[  \left( \bar{q} \sigma_{ij} \tau_k q \right)^2 - 2 \left( \bar{q} \sigma_{0i}  \tau_k q \right)^2 \right] \,,  \\[1ex]\nonumber 
\mathcal{L}_{T^{\text{adj}}_{\bot - \parallel, \mathbbm{1} + \textrm{ISO}}} &=&  \sum_{k=1}^3 \left[  \left( \bar{q} \sigma_{ij} \tau_k T^a q \right)^2 \!-\!  2 \left( \bar{q} \sigma_{0i}  \tau_k T^a q \right)^2 \right]\nn\\
&& \quad + \left( \bar{q} \sigma_{ij} T^a q \right)^2 - 2 \left( \bar{q} \sigma_{0i} T^a  q \right)^2   \,. 
\ee
Here, $\sigma_{\mu\nu}=({\rm i}/4)[\gamma_{\mu},\gamma_{\nu}]$, the $T^a$'s are the generators of $SU(\Nc)$, and the~$\tau_k$'s are related to the generators of the $SU(3)$ flavour space in the same way as the Pauli matrices are related to the generators of~$SU(2)$. The indices~$l$ and~$s$ in these expressions refer to the subspace of the two light quark flavours and the strange quark, respectively. Moreover, 
\be\nonumber 
\mathbbm{1}^{l} \otimes \mathbbm{1}^{l} &=& \!\frac{4}{9} \bigg( \!\mathbbm{1} \otimes \mathbbm{1} +\frac{\sqrt{3} }{2}\!  \left[ \mathbbm{1} \otimes \tau_8 + \tau_8 \otimes \mathbbm{1} \right] + \frac{3}{4}  \tau_8 \otimes \tau_8 \bigg)\,,
\ee
and
\be
\mathbbm{1}^{s} \otimes \mathbbm{1}^{s} &=& \!\frac{1}{3} \mathbbm{1} \otimes \mathbbm{1} - 
\frac{1}{2} \mathbbm{1}^{l} \otimes \mathbbm{1}^{l}  +\frac{2}{3} \; \tau_8 \otimes \tau_8\,.\nn
\ee
In the case of broken $U(1)_{\rm A}$ symmetry, we have 6 additional channels, 
{\allowdisplaybreaks
\be\nonumber 
	\mathcal{L}_{(S+P)_{\mathbbm{1}^l-\textrm{ISO}}} &=& \left( \bar{q}  \mathbbm{1}^l q \right)^2 + \left( \bar{q} \gamma_5 \mathbbm{1}^l q \right)^2 \nn \\[1ex]\nonumber 
	&&\quad - \sum_{k=1}^3 \left[ \left( \bar{q} \gamma_5 \tau_k q \right)^2 + \left( \bar{q} \tau_k q \right)^2 \right]\,,   \\[1ex]\nonumber 
\mathcal{L}_{(S+P)^{\text{adj}}_{\mathbbm{1}^l-\textrm{ISO}}} &=& \left( \bar{q} \mathbbm{1}^l T^a q \right)^2 + \left( \bar{q} \gamma_5 \mathbbm{1}^l T^a q \right)^2  \\[1ex]\nonumber 
&&\quad  - \sum_{k=1}^3 \left[ \left( \bar{q} \gamma_5 \tau_k T^a q \right)^2 + \left( \bar{q} \tau_k T^a q \right)^2 \right]\,,  \nn \\[1ex]\nonumber 
\mathcal{L}_{(S+P)_{s}} &=& \left( \bar{q}  \mathbbm{1}^{s} q \right)^2 + \left( \bar{q} \gamma_5  \mathbbm{1}^{s} q \right)^2 \,,  \nn \\[1ex]\nonumber 
\mathcal{L}_{(S+P)^{\text{adj}}_{s}} &=& \left( \bar{q}  T^a \mathbbm{1}^{s} q \right)^2 + \left( \bar{q}  \gamma_5 T^a \mathbbm{1}^{s} q \right)^2\,,    \nn \\[1ex]\nonumber 
\mathcal{L}_{T_{\bot + \parallel, \mathbbm{1} - \textrm{ISO}}} &=& \left( \bar{q} \sigma_{ij} q \right)^2 + 2 \left( \bar{q} \sigma_{0i}  q \right)^2    \nn  \\[1ex]\nonumber 
&&\qquad - \sum_{k=1}^3 \left[  \left( \bar{q} \sigma_{ij} \tau_k q \right)^2 + 2 \left( \bar{q} \sigma_{0i}  \tau_k q \right)^2 \right] \,,    \\[1ex]\nonumber 
\mathcal{L}_{T^{\text{adj}}_{\bot + \parallel, \mathbbm{1} - \textrm{ISO}}} &=&- \sum_{k=1}^3 \left[  \left( \bar{q} \sigma_{ij} \tau_k T^a q \right)^2 + 2 \left( \bar{q} \sigma_{0i}  \tau_k T^a q \right)^2 \right]\nn\\
&& \quad\quad + \left( \bar{q} \sigma_{ij} T^a q \right)^2 + 2 \left( \bar{q} \sigma_{0i} T^a  q \right)^2   \,. 
\ee
Thus, in total, we end up with 32 four-quark interaction channels in case of QCD with two massless and one heavy quark flavour. Finally, we add that we can relate this basis to the basis underlying our studies of two-flavour QCD. For example, the associated scalar-pseudoscalar and diquark channel are given by
\be
\mathcal{L}_{(\sigma^l - \pi)} &=& - \mathcal{L}_{(V+A)_{\parallel,l}^\text{adj}} - \mathcal{L}_{(V+A)_{\perp,l}^\text{adj}} - \frac{1}{2 N_c} \mathcal{L}_{(V+A)_{\parallel,l}} \nn \\[1ex]
&& \quad - \frac{1}{2 N_c} \mathcal{L}_{(V+A)_{\perp,l}} + \frac 1 2 \mathcal{L}_{(S+P)_{\mathbbm{1}^l-\textrm{ISO}}} \,,
\ee
and
\be
\mathcal{L}_{\text{csc}} &=& \, \mathcal{L}_{(S+P)_{\mathbbm{1}^l-\textrm{ISO}}^\text{adj}} + \mathcal{L}_{(V-A)_{\parallel,l}^\text{adj}} + \mathcal{L}_{(V-A)_{\perp,l}^\text{adj}} \nn \\\nonumber 
&& \qquad - \frac{N_c -1 }{2N_c} \bigg( \mathcal{L}_{(S+P)_{\mathbbm{1}^l-\textrm{ISO}}} + \mathcal{L}_{(V-A)_{\parallel,l}} \nn\\
&& \qquad\qquad\qquad\qquad + \mathcal{L}_{(V-A)_{\perp,l}} \bigg)\,, 
\ee
respectively. The use of these channels may be more convenient in phenomenological applications.



%
\bibliography{qcd}

\end{document}